\pdfoutput=1

\documentclass[manuscript,screen]{acmart}

\AtBeginDocument{%
  \providecommand\BibTeX{{%
    \normalfont B\kern-0.5em{\scshape i\kern-0.25em b}\kern-0.8em\TeX}}}

\setcopyright{acmcopyright}
\copyrightyear{2023}
\acmYear{2023}
\acmDOI{XXXXXXX.XXXXXXX}

\acmConference[TOSEM]{}{2023}{}
\acmPrice{15.00}
\acmISBN{978-1-4503-XXXX-X/18/06}

\usepackage{amssymb}
\usepackage{tcolorbox}
\usepackage{xcolor}
\usepackage{booktabs}
\usepackage[linesnumbered,ruled]{algorithm2e}
\usepackage{graphicx}
\usepackage{multirow}
\usepackage{amsmath}
\usepackage{subfigure}
\usepackage{svg}
\usepackage{geometry}
\makeatletter

\newcommand{\Rmnum}[1]{\expandafter\@slowromancap\romannumeral #1@}
\makeatother

\begin{document}

\title{LibAM: An Area Matching Framework for Detecting Third-party Libraries in Binaries}

\author{Siyuan Li}
\authornote{Both authors contributed equally to this research.}
\email{lisiyuan@iie.ac.cn}
\author{Yongpan Wang}
\authornotemark[1]
\email{wangyongpan@iie.ac.cn}
\author{Chaopeng Dong}
\email{dongchaopeng@iie.ac.cn}
\author{Shouguo Yang}
\email{yangshouguo@iie.ac.cn}
\author{Hong Li}
\authornote{Corresponding author}
\email{lihong@iie.ac.cn}
\author{Hao Sun}
\email{sunhao@iie.ac.cn}
\author{Zhe Lang}
\email{langzhe@iie.ac.cn}
\author{Zuxin Chen}
\email{chenzuxin@iie.ac.cn}
\author{Weijie Wang}
\email{wangweijie@iie.ac.cn}
\author{Hongsong Zhu}
\email{zhuhongsong@iie.ac.cn}
\author{Limin Sun}
\email{sunlimin@iie.ac.cn}
\affiliation{%
  \institution{School of Cyber Security, University of Chinese Academy of Sciences and Institute of Information Engineering, Chinese Academy of Sciences}
  \city{BeiJing}
  \country{China}
}

\renewcommand{\shortauthors}{Siyuan Li and Yongpan Wang, et al.}

\begin{abstract}
Third-party libraries (TPLs) are extensively utilized by developers to expedite the software development process and incorporate external functionalities. Nevertheless, insecure TPL reuse can lead to significant security risks. Existing methods, which involve extracting strings or conducting function matching, are employed to determine the presence of TPL code in the target binary. However, these methods often yield unsatisfactory results due to the recurrence of strings and the presence of numerous similar non-homologous functions. Furthermore, the variation in C/C++ binaries across different optimization options and architectures exacerbates the problem. Additionally, existing approaches struggle to identify specific pieces of reused code in the target binary, complicating the detection of complex reuse relationships and impeding downstream tasks. And we call this issue the poor interpretability of TPL detection results.

In this paper, we observe that TPL reuse typically involves not just isolated functions but also areas encompassing several adjacent functions on the Function Call Graph (FCG). We introduce LibAM, a novel Area Matching framework that connects isolated functions into function areas on FCG and detects TPLs by comparing the similarity of these function areas, significantly mitigating the impact of different optimization options and architectures. Furthermore, LibAM is the first approach capable of detecting the exact reuse areas on FCG and offering substantial benefits for downstream tasks. To validate our approach, we compile the first TPL detection dataset for C/C++ binaries across various optimization options and architectures. Experimental results demonstrate that LibAM outperforms all existing TPL detection methods and provides interpretable evidence for TPL detection results by identifying exact reuse areas. We also evaluate LibAM's scalability on large-scale, real-world binaries in IoT firmware and generate a list of potential vulnerabilities for these devices. Our experiments indicate that Area Matching framework performs exceptionally well in the TPL detection task and holds promise for other binary similarity analysis tasks. Last but not least, by analyzing the detection results of IoT firmware, we make several interesting findings, for instance, different target binaries always tend to reuse the same code area of TPL. The datasets and source code used in this paper are available at \href{https://github.com/Siyuan-Li201/LibAM}{https://github.com/Siyuan-Li201/LibAM}.
\end{abstract}



\thanks{This work was partially supported by the National Key Research and Development Program of China (2022YFB3103904), the National Natural Science Youth Foundation (62002342), and the National Natural Science Foundation of China (61931019)}


\maketitle

\section{Introduction}
In order to accelerate the software development process and integrate external functionalities, developers frequently rely on existing code from open-source code repositories and package management platforms such as \textit{Conan} \cite{conan}, \textit{Vcpkg} \cite{vcpkg}, \textit{GitHub} \cite{github}, and \textit{SourceForge} \cite{sourceforge}. These resources are referred to as third-party libraries (TPLs) \cite{javatplreview}. As the open-source software (OSS) ecosystem continues to expand, an ever-increasing number of software projects are being constructed utilizing TPLs as their foundation. A recent report from Synopsys \cite{Blackduck} indicates that a staggering 97\% of audited software incorporates at least one TPL.

However, the reliability of a large number of TPLs is difficult to guarantee, and security vulnerabilities in one software project can easily propagate throughout the software supply chain, thereby impacting other software projects. Among the audited software in Synopsys report \cite{Blackduck}, 81\% contain at least one known security vulnerability. Moreover, developers who unintentionally introduce improper TPLs may also violate open-source licensing regulations, resulting in legal complications. For instance, both Cisco and VMware have encountered significant legal issues due to non-compliance with the stipulations set forth in the Linux license \cite{Cisco, VMware}. This underlines the importance of being vigilant when incorporating TPLs into software projects, in order to maintain both the security and legal integrity of the resulting applications.

Generally speaking, TPL detection is a generic technology that detects code reuse relationships between software and can be applied to a large number of downstream tasks. On the one hand, both developers and users are keen to identify and manage TPLs in their software to mitigate security risks associated with TPL reuse \cite{javatplreview}. Moreover, the detection of software plagiarism and open-source code infringement has gained significant attention from researchers in recent years \cite{OSSPolice}. On the other hand, TPL detection results can be utilized for 1-day vulnerability detection and malware identification. For instance, Firmsec \cite{FirmSec} employed TPL detection technology to uncover numerous 1-day vulnerabilities in IoT firmware, demonstrating that the actual impact scope of known vulnerabilities detected via TPL detection technology often extends far beyond what is reported in the CVE \cite{CVE} or NVD \cite{NVD} databases. This paper focuses on the study of the generic TPL detection technique and evaluates the application of the vulnerability association task.

In an effort to counteract the potential risks posed by unreliable TPLs, numerous researchers have focused on the development of effective TPL detection approaches for software applications. Originally, the majority of research efforts concentrated on TPL detection in Java \cite{java1,java2,java3,java4,java5,java6,java7}. Recently, there has been a growing interest in TPL detection within C/C++ binaries \cite{BAT, OSSPolice, B2SFinder, ISRD, LibDX, LibDB, CENTRIS}. TPL detection in C/C++ binaries presents even greater challenges, as binaries compiled using diverse optimization options and architectures exhibit significant differences \cite{Code_Similarity_survey1, Code_Similarity_survey2}. In detail, Current approaches for TPL detection in C/C++ binaries typically involve gathering an extensive database of candidate TPLs and subsequently determining which of these have been reused in the target binary. These approaches can be broadly classified into two main categories: constant-based works \cite{BAT, OSSPolice, B2SFinder, LibDX} and function similarity-based works \cite{ISRD, LibDB, ModX}. Constant-based works identify TPL reuse within the target binary by extracting identical constant features, such as strings \cite{BAT, LibDX}, function names \cite{OSSPolice}, and jump tables \cite{B2SFinder}, from both the target binary and the candidate TPLs. In contrast, function similarity-based works involve comparing all the functions of the target binary with those of the candidate TPLs. Subsequently, a predetermined threshold is set to establish whether reuse has occurred. ISRD \cite{ISRD} addresses reuse detection by identifying more than half of the similar functions in TPL, while LibDB \cite{LibDB} depends on over three connected functions on FCG to ascertain reuse. ModX \cite{ModX} divides functions with the same functionality into a group by clustering and sets a threshold for every group.


However, it is essential to note that existing TPL detection technologies for C/C++ binary exhibit certain limitations. These limitations may hinder their performance and scalability in addressing the various downstream tasks for which they are intended. We summarize the limitations of the existing works in the following three points:


Firstly, existing third-party library (TPL) detection approaches are difficult to achieve both high accuracy and high robustness across different optimization options and architectures. On the one hand, The features selected by existing methods are not always working. Although constant-based approaches, which rely on strings \cite{BAT, LibDX}, function names \cite{OSSPolice}, and jump tables \cite{B2SFinder}, exhibit robustness in varying optimization options and architectures, these approaches may falter when the number of constants is limited or when repeated constants appear in distinct binaries, leading to reduced performance in detecting TPLs. In contrast, function similarity-based approaches can detect every instance of reuse by comparing all functions \cite{ISRD, LibDB, ModX}. 
Unfortunately, the accuracy of isolated function matching significantly declines as the number of similar functions increases and variations across different optimization options and architectures, compromising the accuracy of the TPL detection results. On the other hand, the detection granularity of existing approaches does not match the reuse granularity. Current techniques generally set a threshold value for the entire binary \cite{LibDB, ModX} or source file \cite{B2SFinder, OSSPolice}, and when the number of matched features reaches this threshold, the entire binary or source file is considered to be reused. However, In numerous cases, software reuses only a portion of TPLs (Partial Reuse) \cite{OSSPolice}. As a result, some small-scale reuse instances may be missed because the threshold is not met, and some libraries with a large number of similar features may be mistakenly reported, causing false negatives and false positives, which are described in detail in Section 2.2.


Secondly, the TPL detection results of current approaches are limited to the file-level granularity, which is insufficient for uncovering complex reuse relationships and interferes with downstream tasks. Complex reuse relationships may include partial reuse and pseudo-propagation reuse (the relation between A and B in which both A and B reuse C) which is proposed in the previous paper \cite{B2SFinder}. Existing approaches can only detect which TPLs are reused by the target software, but cannot further detect which part of the code in TPLs is actually reused. These coarse-grained detection results are not applicable to the need for fine-grained results for downstream tasks. For instance, ModX \cite{ModX} demonstrates that a large number of vulnerabilities are often concentrated in a small portion of the software code. Users want to identify which specific parts of the TPLs are reused by the software for refined software management or to assess whether sensitive portions (vulnerabilities or malware) of the TPLs have been reused. Developing and refining such methods would significantly enhance the effectiveness of TPL detection and reduce the risks associated with the undetected reuse of vulnerabilities or malware. 

Thirdly, the datasets used in previous research for TPL detection tasks are in a single compiled environment \cite{ISRD, LibDB}, limiting the evaluation of the robustness and scalability in real scenarios. Even though existing works have gathered a substantial number of real-world binaries and manually labeled the ground truth to analyze their accuracy, the TPL detection datasets are limited to default optimization options and a single architecture. In many situations, target binaries are compiled from varying architectures and optimization options (e.g., binaries in firmware), but existing works \cite{ISRD, LibDB} only evaluate the accuracy of the function similarity matching task rather than TPL detection task under different compilation environments. The absence of diverse datasets increased the difficulty of appraising the scalability of TPL detection approaches. To tackle this issue, it is crucial to develop new datasets that encompass a broader range of optimization options and architectures, capturing the inherent variability and complexity of modern software systems.

In this paper, we introduce LibAM, a novel Area Matching framework that exhibits accuracy and robustness across various optimization options and architectures and can detect exact reuse areas. We find that if one function is reused, its callee functions are also reused. Therefore, different from existing works that rely on function granularity matching, we leverage the function call relationships to connect isolated functions into areas on FCG and compare the similarity of these areas rather than single functions to judge reuse. Function inlining, function call deletion, and other changes from the compiler that are fatal to isolated function matching, have little effect on the graph similarity of FCGs. There are two primary modules in LibAM: Area Generation and Area Comparison. This innovative framework can be effectively employed for two tasks: TPL detection and Reuse area detection. The former aims to detect which TPLs are reused by the target binary, and the latter aims to detect which functions (exact reuse areas) of the target binary are from TPLs. LibAM takes target binaries and TPL binaries as inputs, and finally outputs which TPLs are reused in target binaries and a list of reused functions. Therefore, LibAM can provide specific reuse areas for each TPL detection result, thus providing interpretability and being used to detect complex reuse relationships.

In detail, LibAM commences with conducting a comprehensive comparison of all target functions and TPL functions through a vector-searching technique. Subsequently, the anchor extension phase enables the connection of previously isolated function nodes into areas on the Function Call Graph (FCG). Finally, LibAM assesses whether a particular area has been reused by calculating the similarity of function areas. LibAM outputs candidate TPLs for the TPL detection task while generating a reused function name list for the Reuse area detection task. Note that the novel area detection task can help analyze complex reuse relationships, and we obtain two interesting findings based on this in Section 4.8. In addition, we show that by detecting exact reuse areas, vulnerabilities can be associated by matching vulnerable functions rather than by matching vulnerable TPLs, which filters false positives for vulnerability associations caused by partial reuse.

We evaluate LibAM and the state-of-the-art (SOTA) works including LibDX, B2SFinder, ISRD and LibDB using the public dataset from a previous paper \cite{OSSPolice} as well as the first dataset for different optimization options and architectures built by ourselves. Our experiments show that LibAM outperforms all existing TPL detection works and beats the SOTA work \cite{LibDX} by an average of 24\% in precision and 36\% in recall even across different optimization options and architectures. 
Moreover, compared to previous methods, which failed in the exact reuse area detection task, LibAM has demonstrated the feasibility of doing so with 0.99 precision and 0.844 recall. We further evaluate the ability of area matching to detect the exact reuse area for complicated reuse relationships and downstream tasks by detecting reuse relationships in large-scale IoT firmware and associating vulnerabilities introduced due to TPL. ModX \cite{ModX} shows that a large number of vulnerabilities are often in a small part of the software code. By detecting specific reuse areas, we can determine whether vulnerable functions are reused, thus avoiding false vulnerability associations due to file-level reuse detection.

We summarize our main contributions below:
\begin{itemize}
\item We propose a novel framework LibAM, which represents a significant improvement over all existing TPL detection works both in the public real-world dataset and in different optimization options or architectures.

\item To the best of our knowledge, this is the first work to attempt to detect the exact reuse area which is beneficial for downstream tasks. 

\item We build the first dataset in different optimization options and architectures for TPL detection, which allows us to evaluate the robustness of existing works.

\item We evaluated the scalability of LibAM using a large-scale real-world IoT firmware dataset and generate potential firmware vulnerabilities to show one application of LibAM. 
\end{itemize}

The remainder of the article is as follows: Section 2 introduces the problem description, motivation and notations of this paper. Section 3 describes the design details of LibAM. Section 4 compares LibAM with existing TPL detection techniques. Section 5 discusses the results and limitations of LibAM. Section 6 presents related work.

\begin{figure*}
\centering
    \includegraphics[scale=.36]{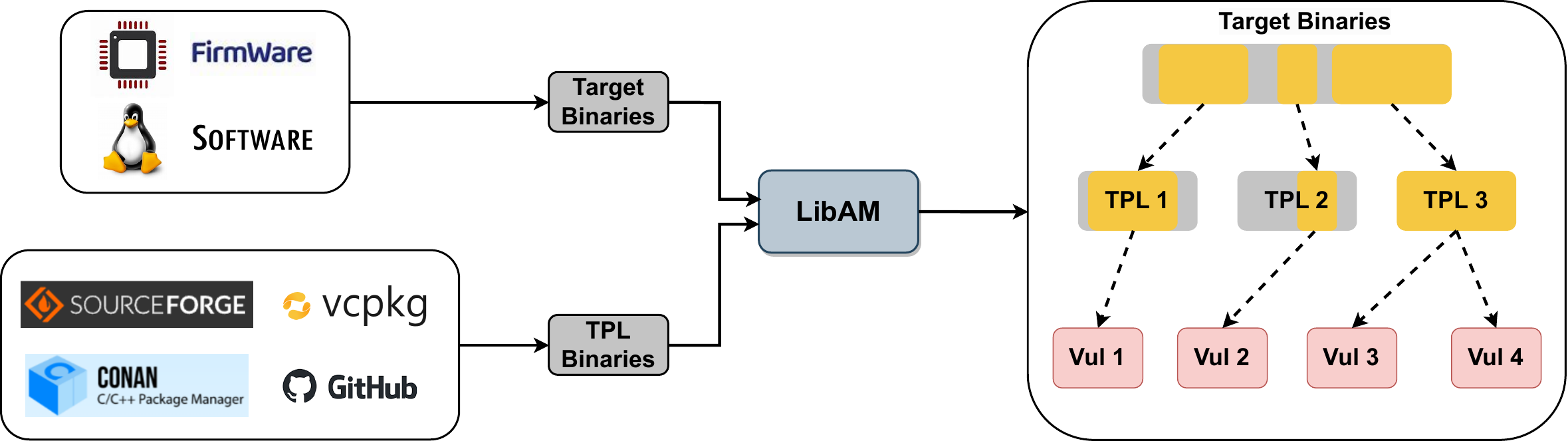}

\caption{Application scenarios of LibAM. LibAM takes target binaries and TPL binaries as inputs. Target binaries are extracted from firmware and software, while TPL binaries are collected from package management platforms or manually compiled. The output of LibAM is the TPLs and reuse areas of the target binaries (in yellow in the figure), which can be further associated with vulnerabilities.}
\vspace{-1.0em}
\end{figure*}

\section{Preliminary}
\subsection{Problem Description}
TPL detection takes the target binaries and widely used TPL binaries as input, and attempts to detect which TPLs are reused by the target binary. We aim to design a generic approach that can solve the TPL detection task for C/C+ binaries in multiple scenarios.

Firstly, we want to detect TPLs in target binaries that come from both software and firmware as in Figure 1. Binaries in software tend to be compiled in a single architecture (e.g. B2SFinder \cite{B2SFinder} only detects PE files under Windows). However, binaries in firmware tend to come from different architectures and raise the demands for TPL detection technology. Note that both binaries in software and firmware are stripped and we can't extract function names by traditional tools like \textit{nm}. Our proposed method is designed to address these challenges by enabling the detection of TPLs in target binaries from various optimization and architectures.

Secondly, our approach aims to detect TPLs that are reused through dynamic links, static links, and direct copies. For dynamic links, target software or firmware retains dynamic link library files, such as .dll or .so files, within the file directory. We build a TPL database consisting of commonly used dynamic link library files and utilize the dynamic link library files in the target software or firmware as the target binary for detection purposes. For static links or direct copies, intricate reuse relationships present challenges for TPL detection. To overcome this, we gather binaries from prevalent C/C++ projects to create the TPL database, subsequently detecting which sections of code in the target binaries originate from these TPL binaries. It is worth noting that these target binaries may make minor modifications to the reused TPL, for example, some binaries slightly modify the reused code to remove string-print instructions without changing the semantics. As in the case of \textit{minizip}, which removes the string-print instructions in the \textit{$BZ2\_BlockSort$} function of \textit{bzip2}. We aim to identify similar areas from massively different codes and overcome minor modification differences. For drastic changes that severely affect the semantics, it is a serious challenge whether code with large changes is considered reused and how to recognize them even for humans, which is outside our dataset and our goals. We only consider minor reusable code changes in the public dataset from previous work \cite{ISRD}.

Thirdly, we aim to detect both reused TPLs and reused code areas. Prior work on TPL detection could not ascertain which portions of code in TPLs were reused, a limitation we term as poor interpretability of TPL results. Research \cite{OSSPolice, B2SFinder} has demonstrated that a large number of TPL reuses are partial, and file-level TPL detection results cannot determine if sensitive parts (e.g., vulnerable code or malware) are reused. This limitation obstructs numerous downstream tasks, including 1-day vulnerability correlation, malware detection, software plagiarism detection, and more. Our approach tends to identify which sections of code in the target binary reuse specific parts of the code in the TPL, providing interpretable evidence for the TPL detection results. Additionally, since function names in TPL binaries are accessible, by gathering patch information from the NVD, we can detect if vulnerable functions are present in the reuse areas by comparing the reused TPL function names with vulnerability names, even when the function names in target binaries are stripped. For the TPL detection task, we collect widely reused TPL binaries and determine which of them are reused. For the Area detection task, we pinpoint which functions of the target binaries originate from TPLs, and we can obtain the reused function names since the corresponding function names in TPLs are readily available.


\begin{figure}
    \includegraphics[scale=.42]{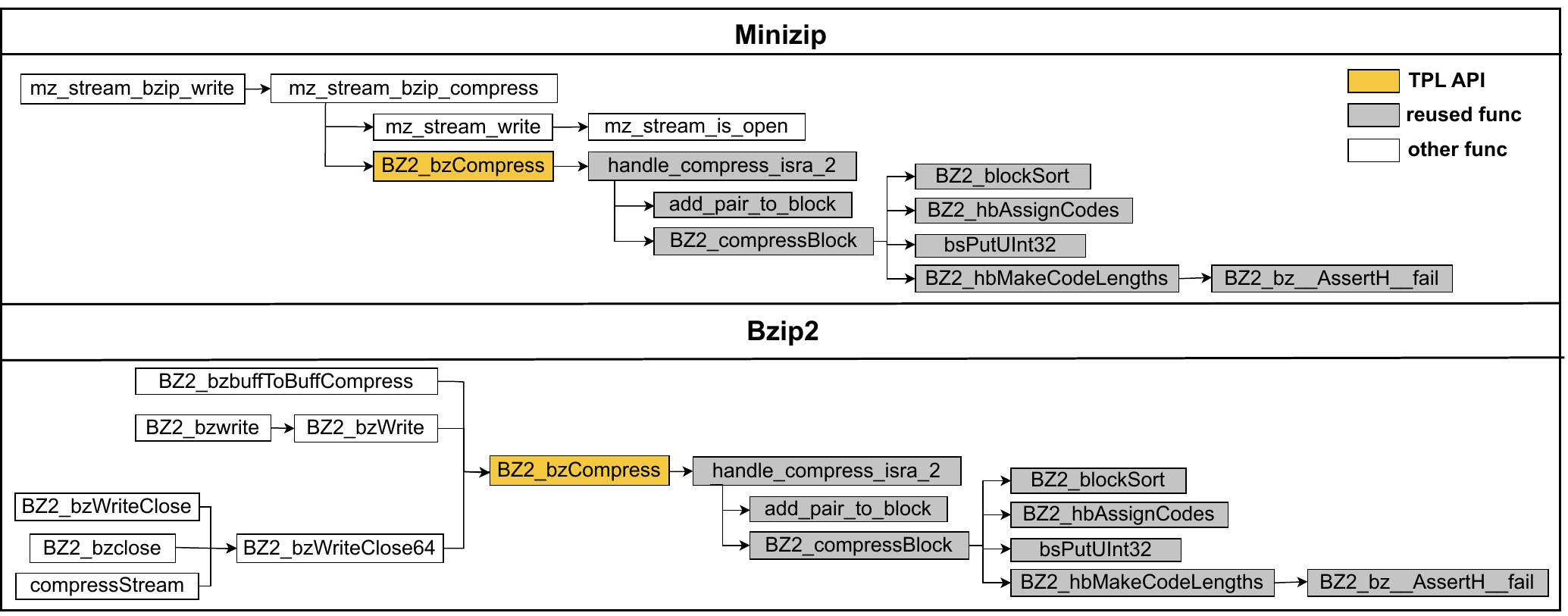}

\caption{A motivation example for LibAM. The upper part depicts a portion of the FCG for \textit{Minizip}, while the lower part presents a portion of the FCG for \textit{Bzip2}. Each block represents a function, with yellow blocks indicating the reused TPL API functions and gray blocks denoting functions that are reused alongside the API functions.}
\vspace{-1.0em}
\end{figure}

\subsection{Motivation}
In light of the limitations of existing third-party library (TPL) detection methods, our primary objective is to develop a novel TPL detection framework that demonstrates accuracy and robustness across a diverse range of optimization options and architectures. Furthermore, our proposed approach aims to detect exact reuse areas, enabling more precise and fine-grained analysis of TPL reuse in software systems. To achieve this, we draw upon insights from the patterns of function reuse observed in real-world software applications in ISRD \cite{ISRD}. 

We have found that when a particular function is reused, its callee functions are also frequently reused. As illustrated in Figure 2, for example, since \textit{Minizip} reuses the function \textit{BZ2\_bzCompress} from \textit{Bzip2}, the callee functions of \textit{BZ2\_bzCompress} are also reused. This observation suggests that leveraging function call relationships may provide a more accurate and robust approach to TPL detection compared to existing methods, which primarily rely on isolated function granularity matching.

Besides, there are many partial reuses in the actual scenario, and the reuse proportion is much smaller than file-level TPL. In Figure 3, due to the fact that \textit{snkfile2k} is mostly reused as test binaries within the project, we have shown the reuse proportion of 41 reuse relationships in Dataset\_ISRD (A TPL dataset from previous work \cite{ISRD}, which is described in detail in Section 4.1), except for \textit{snkfile2k} related reuse. It can be seen that over half of the reuse relationships only reuse less than half of the functions in TPLs, and these partial reuses may have a serious impact on the TPL detection results obtained by matching isolated functions.

Building on these insights, we propose a novel approach that connects isolated functions into areas on the Function Call Graph (FCG) based on their call relationships and compares the similarity of these areas, rather than single functions, to determine reuse. Unlike LibDB \cite{LibDB} which uses a simple rule that TPL with three connected functions on FCG are reused, we want to compare the structural similarity and node alignment of the function areas on FCG. We found an overlapping phenomenon in LibDB, which is described in detail in Section 3.2.2, and we propose an Anchor Alignment algorithm to solve the overlapping problem of LibDB by comparing the structural similarity of areas through GNN and subjected to the RANSAC algorithm \cite{RANSAC} in the field of image alignment to generate a one-to-one correspondence of area node correspondence.

\begin{figure}
    \includegraphics[scale=.28]{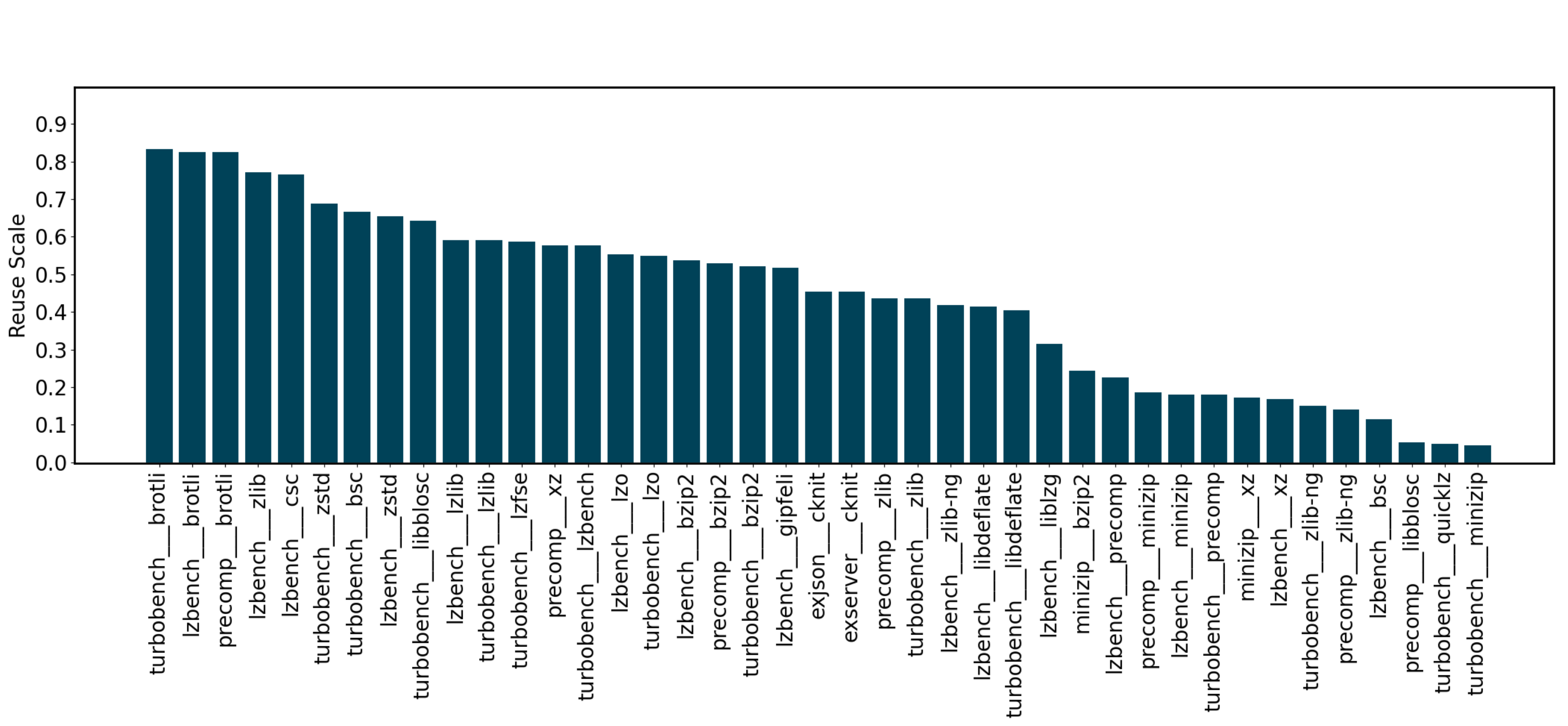}

\caption{Reuse proportion in Dataset\_ISRD. We manually analyzed the reuse areas of each reuse relationship in Dataset\_ISRD and counted the number of reused TPL functions and the total number of TPL functions. The quotient of the two values is the reuse proportion.}
\end{figure}

Moreover, LibDB \cite{LibDB} shows that there may be many false negatives and false positives during the function-matching phase. However, the Area Matching framework can effectively revise these mistakes, thus mitigating the impact of varying optimization options and architectures on TPL detection accuracy. What's more, we seek to evaluate the ability of area matching to detect the exact reuse area for complex reuse relationships and its implications for downstream tasks, such as vulnerability analysis and software management. For instance, ModX \cite{ModX} demonstrates that a large number of vulnerabilities are often concentrated in a small portion of the software code. By detecting specific reuse areas, our approach can determine whether vulnerable functions have been reused, thereby avoiding false vulnerability associations that may arise from file-level reuse detection.

\subsection{Definition}
To ensure a more coherent presentation and facilitate comprehension, this section focuses on standardizing the terminology employed throughout this paper. The detected binary extracted from software or firmware is called \textit{\textbf{target binary}} while the binary in the TPL database is denoted as the \textit{\textbf{TPL binary}}. Additionally, the Control Flow Graph is abbreviated as \textit{\textbf{CFG}} and the Function Call Graph as \textit{\textbf{FCG}}. Furthermore, the Attributed Control Flow Graph, which is a concept leveraged by Gemini \cite{Gemini}, is designated as \textit{\textbf{ACFG}}. Finally, we define the \textbf{\textit{Anchor}} and \textbf{\textit{Reuse Area}} to facilitate a clearer explanation of the algorithm.

\subsubsection{\textbf{Anchor}}

In the process of generating and comparing areas, our initial step involves matching functions in the target binary and the TPL binaries by employing a function similarity calculation method. Functions that are successfully matched are termed \textit{\textbf{anchors}}, while a pair of matched functions (one originating from the target binary and the other from the TPL binary) are referred to as an \textit{\textbf{anchor pair}}. As illustrated in Figure 4, the gray nodes on both the target FCG and TPL FCG, following the Anchor Detection phase, signify anchors.

Formally, let \( \mathcal{F}_{\text{target}} \) represent the set of functions extracted from the target binary, and \( \mathcal{F}_{\text{TPL}} \) denote the set of functions from the TPL binary. The function similarity calculation method is denoted as \( \text{Sim}(f_i, f_j) \), where \( f_i \in \mathcal{F}_{\text{target}} \), \( f_j \in \mathcal{F}_{\text{TPL}} \). An anchor \( a \) is a function in \( \mathcal{F}_{\text{target}} \) that successfully matches with at least one function \( b \) in \( \mathcal{F}_{\text{TPL}} \), according to the similarity threshold \( \theta_s \):

\begin{equation}
a \in \mathcal{F}_{\text{target}}, \quad \exists b \in \mathcal{F}_{\text{TPL}} : \text{Sim}(a, b) \geq \theta_s
\end{equation}

An anchor pair \( (a, b) \) is formed by an anchor \( a \) and its matching function \( b \) from \( \mathcal{F}_{\text{TPL}} \).

\subsubsection{\textbf{Area}}

We connect the isolated anchors and all their sub-functions on FCG to generate a function list which we term the \textit{\textbf{function area}}. In essence, the function area is a node list that encompasses all nodes reachable from a selected anchor node on the FCG. It is essential to note that we generate two distinct areas for each anchor pair: one being the \textit{\textbf{target area}} and the other the \textit{\textbf{TPL area}}. 

Given an anchor pair \( (a, b) \), the target area \( \mathcal{A}_{(a)} \) and candidate area \( \mathcal{A}_{(b)} \) are formally defined as:

\begin{equation}
\mathcal{A}_{(a)} = \{ f \in \mathcal{F}_{\text{target}} \, | \, \text{there exists a path from } a \text{ to } f \text{ in } \text{FCG}_{\text{target}} \}
\end{equation}

\begin{equation}
\mathcal{A}_{(b)} = \{ f \in \mathcal{F}_{\text{candidate}} \, | \, \text{there exists a path from } b \text{ to } f \text{ in } \text{FCG}_{\text{candidate}} \} 
\end{equation}

Our approach compares function areas, rather than isolated functions, to detect TPLs more effectively. The final list of detected reused functions is referred to as the \textit{\textbf{reuse area}}. In Figure 4, the functions enclosed within the dashed box constitute the areas.


\begin{figure*}
\centering
    \includegraphics[scale=.28]{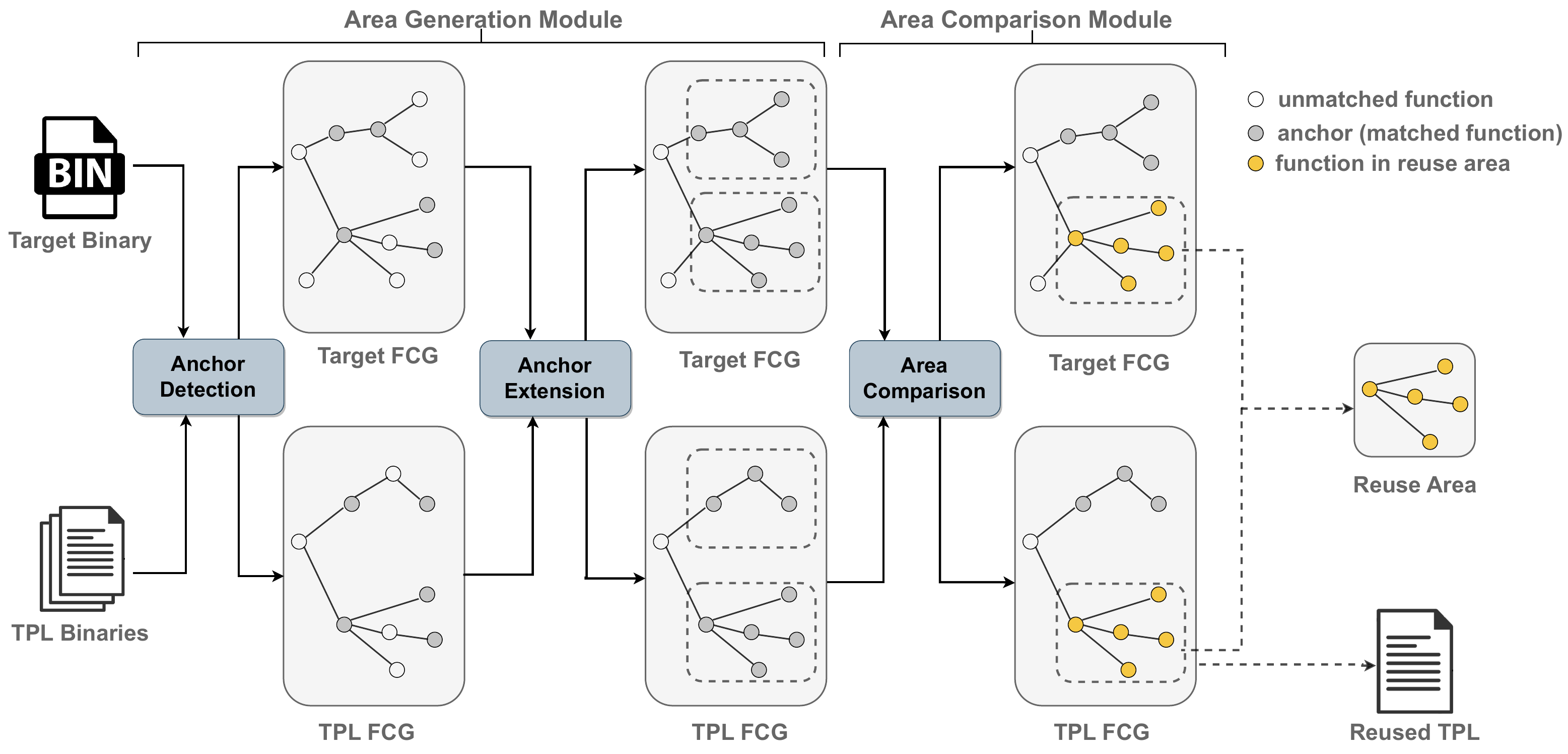}

\caption{The workflow of LibAM. We mark the matched function as gray on the FCG. After the Area Generation module, we represent the areas connected by anchors as dashed boxes. After the Area Comparison module, the reused functions in the correct area are marked as yellow. }
\vspace{-1.0em}
\end{figure*}

\section{Methodology}
\subsection{Overview}
We propose a novel Area Matching framework, LibAM, for detecting both the TPLs and exact reuse areas in the target binary. The workflow of LibAM, presented in Figure 4, consists of two modules: Area Generation  (Section 3.2), and Area Comparison (Section 3.3). We take a target binary and TPL binaries as inputs, and LibAM generates a reused TPL list and a reused function list (reuse area) for the target binary after detection.

In the Area Generation module, we create several areas for both the target binary and TPL binaries. The initial step involves extracting functions and FCGs from the target binary and TPL binaries using IDA Pro \cite{IDA_Pro}. Then we conduct a comparison of the functions. Functions that successfully matched are considered anchors. To expedite the function-matching process, we employ annoy \cite{annoy}, a high-speed vector searching engine. Subsequently, we establish connections between anchors and their callee functions on FCG, generating a function area for each anchor. It is important to note that function areas are generated in pairs: one area corresponds to the matched functions within the target binary and the other is associated with the corresponding functions in the TPL binaries.

In the Area Comparison module, we calculate the similarity between each target area and its corresponding TPL area to determine whether it constitutes a reuse area. The similarity is calculated based on two factors: structural similarity and alignment length. We employ Embedded-GNN (Section 3.3.1) to calculate the structural similarity and our Anchor Alignment algorithm (Section 3.3.2) to compute the alignment length. Ultimately, we ascertain whether an area is genuinely reused by considering these two factors.

Finally, after detecting the reuse areas between the target binary and every TPL binary, LibAM generates a reused TPL list for the TPL detection task and a reused function list for the Area detection task. Although the functions in the target binary are without names, LibAM can use the corresponding function names in TPLs to generate the reused function list.


\subsection{Area Generation}

In this section, our target is to identify similar functions between the target binary and the TPL binaries by employing function similarity calculation techniques and subsequently generate areas for these functions. First, we carry out the anchor detection phase to compare the functions. Following this, we conduct an anchor extension phase to generate areas for each of the identified anchors.

\begin{figure*}
\centering
    \includegraphics[scale=.36]{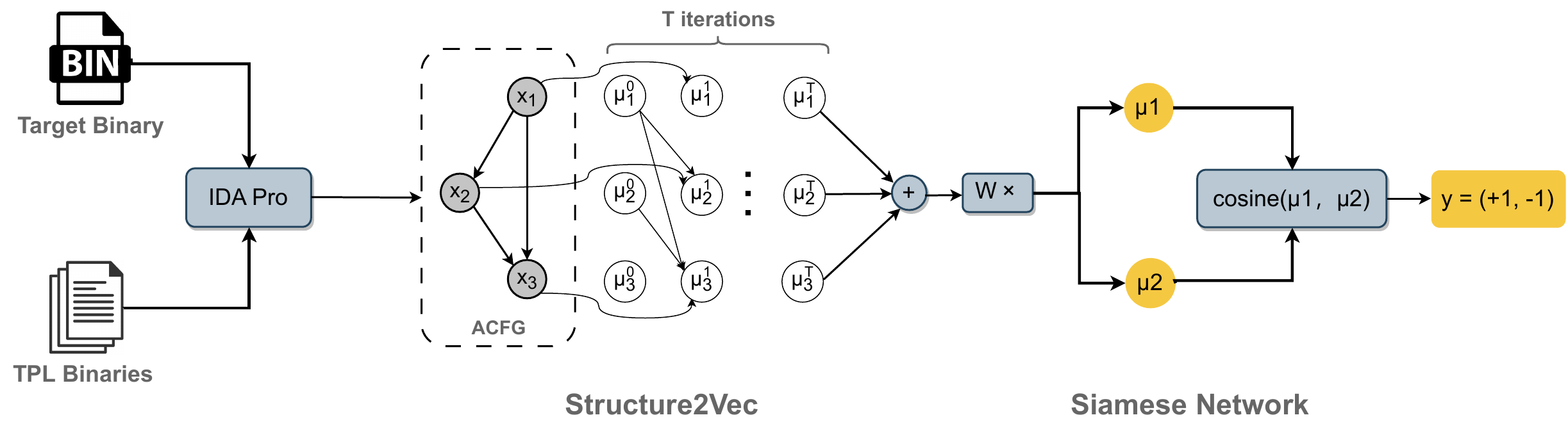}

\caption{Network structure of Structure2Vec. We input the function ACFGs from both target binaries and TPL binaries into the GNN. After T iterations, output vectors are obtained and vector similarity is calculated using cosine similarity. In the figure, the parameters of the two GNNs are shared, forming a Siamese network. }
\vspace{-1.0em}
\end{figure*}

\subsubsection{Anchor Detection}
Existing works \cite{ISRD, LibDB} directly filter these matched functions as reuse detection results. 
Although function similarity matching can distinguish homologous functions from non-homologous functions, its accuracy degrades severely with an increasing number of non-homologous similar functions, as well as in different optimization options and architectures.
Instead, we propose taking the matched functions as anchors and connecting the isolated functions into areas on FCG for further comparison.

As same to previous works \cite{LibDB, ModX}, we choose the improved Gemini in LibDB \cite{LibDB} as our function similarity calculation approach for several reasons. Firstly, previous TPL detection works, such as LibDB \cite{LibDB} and ModX \cite{ModX}, have utilized Gemini for matching tasks, which is a fast and scalable method. Besides, LibAM aims to detect TPLs by comparing areas on FCG and we focus on the improvement brought by area similarity compared to function similarity. Therefore, we leave the enhancement of the new function similarity matching tool itself for future work. Finally, our experiments have demonstrated that with the addition of area comparison framework, Gemini can deliver very good results and it is easy to replace Gemini with new function similarity calculation works to obtain more suitable anchors for specific scenarios in the future.

As in LibDB \cite{LibDB}, we first extract a 7-dimensional vector for each basic block by calculating 7 types of statistical information: the number of the string constants, numeric constants, transfer instructions, call instructions, all instructions, arithmetic Instructions, and offspring numbers. Then, we use each 7-dimensional vector as a node on CFG to transform the function into an Attribute Control Flow Graph (ACFG), which is generated for each function that has block numbers above 5 and instruction numbers above 10 in both the target binary and TPL binaries. Finally, we input the ACFG into the Structure2vec network to obtain the vector representation of the function. The structure of the GNN network is as in Figure 5. We utilize Dataset\_OSS in Section 4.1 to train the Siamese architecture of two Structure2vec networks with shared parameters and optimize it using triplet cosine loss of equation (4). Further details can be found in LibDB \cite{LibDB}.

\begin{equation}
Loss = \frac{1}{m} \sum_{0}^{m}max\left ( \cos\left ( a,n \right ) -\cos \left ( a,p \right ) +\epsilon ,0 \right ) 
\end{equation}

In equation (4), $m$ represents the batch size during training, $cos ( a,p )$ represents the cosine similarity between binary function vectors of different compilation options or architectures that are compiled from the same source code, and $cos ( a,n )$ represents the cosine similarity between function vectors of different source codes. $\epsilon$  is a constant above 0, here we choose 0.2 as in LibDB.

Finally, we calculate the cosine similarity between all function vectors in the target binary and TPLs. Based on our experiment, we set a reasonable threshold of 0.72, which can best distinguish positive samples from negative samples in Dataset\_OSS. When the cosine similarity between two functions exceeds 0.72, we consider these functions as anchors. Besides, we use the \textit{annoy} engine \cite{annoy} to accelerate vector searching. TPL's function embedding process can be performed offline and stores all function vectors in a vector database. For each function of the target binary, we use annoy to retrieve it in the vector database for a fast function comparison. As same in LibDB \cite{LibDB}, we filter top-200 TPLs for each target binary and top-100 TPL functions for each target function to further accelerate the vector searching phase. The details of thresholds are in Section 4.

\subsubsection{Anchor Extension} 
After the anchor detection phase, we generate an anchor pair list, where each anchor pair consists of a function in the target FCG (target anchor) and the corresponding function in the TPL FCG (TPL anchor). In other words, the anchor list represents the matched nodes between the two FCGs. The goal of this phase is to generate an area for each anchor.

The existing works take anchors directly as the result of TPL detection, but the results are not ideal due to a significant number of false positives and false negatives in anchors \cite{LibDB}. To address this issue, LibAM aims to use the function call relationships to link isolated function nodes into areas on FCG. Our observation indicates that when a function is reused in the target binary, its callee functions are reused together. 

\begin{figure*}
\centering
    \includegraphics[scale=.30]{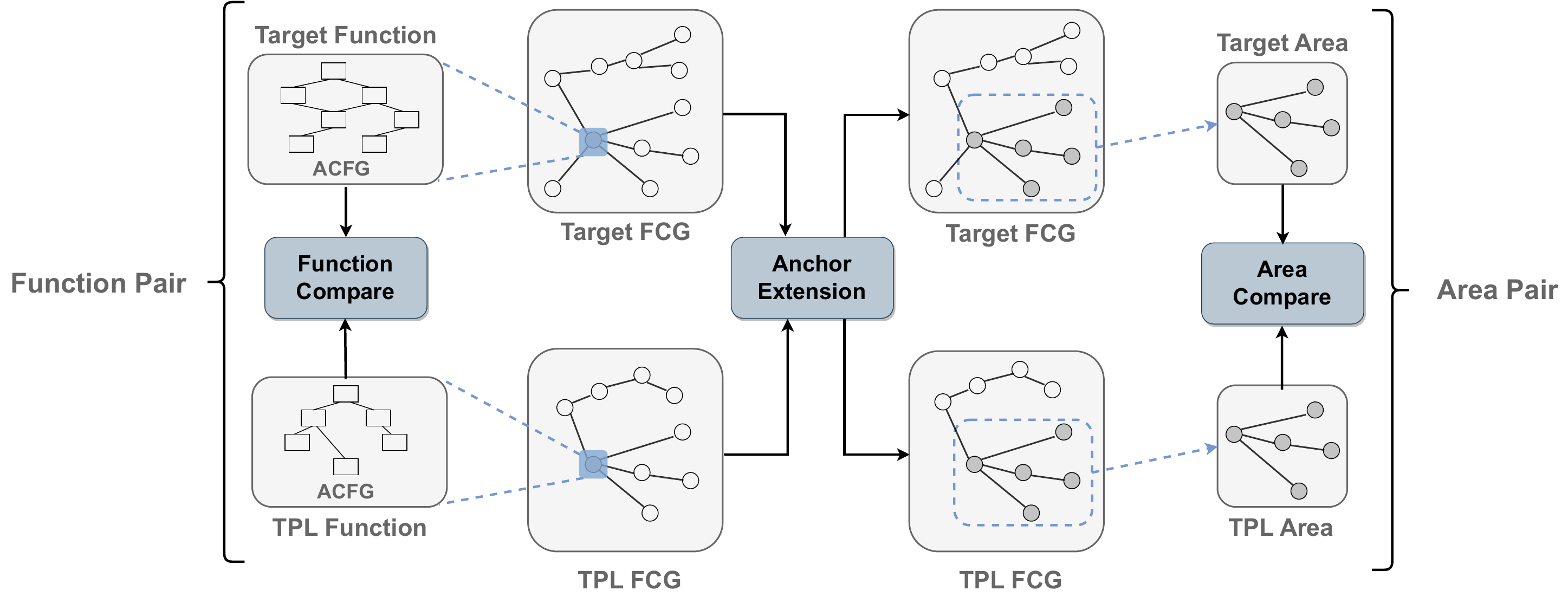}

\caption{Schematic diagram of the Anchor Extension phase. }
\vspace{-1.0em}
\end{figure*}

Based on this insight, in Figure 6, we treat the anchor and all callee functions of it as an area and compare the whole area to detect TPLs. The Area Comparison module takes function pairs as input and generates an area pair including a target area and a TPL area for each function pair. Note that the target area is a function list without function names while the TPL area is a function list with function names. After the Area Comparison module, we can get a reused function name list for the target binary.


\subsection{Area Comparison}
After obtaining two areas of the anchor pair, we aim to compare the similarity of the two areas to determine whether they are reuse areas. We calculate the area similarity by using two factors: the structural similarity $S$ and the alignment length factor $L$. We set two separate thresholds for them and only when both the structural similarity and the alignment length factor exceed the 
thresholds, we consider the areas as actual reuse areas. We will now describe the details of the two factors. The details of thresholds are in Section 4.


\subsubsection{Structural Similarity}
We leverage an Embedded-GNN to generate vectors for areas and regard the vector cosine similarity as structural similarity $S$. The GNN is the same Structure2vec network as in Section 3.2.1. Note that the vector of the function node in the area is just the vector in Anchor Detection, as in Figure 7.

Structure2vec operates on the principle of leveraging graphical model inference methodologies, where features associated with individual vertices, represented by $x_i$, undergo a systematic aggregation process that adheres to the underlying graph topology. After multiple rounds of this recursive procedure, an innovative feature representation (also known as an embedding) is generated for each vertex, effectively capturing the complex interdependencies among vertex features as well as the intrinsic properties of the graph structure. The training procedure can be broadly divided into three main steps: initialization, iterative neighborhood aggregation, and optimization using a supervised learning objective. 

Firstly, we initialize the node embeddings using a node feature matrix. This matrix contains initial feature vectors for all nodes in the graph. The initial embeddings serve as a starting point for the iterative neighborhood aggregation process that follows.

Then, at the heart of Structure2Vec lies the iterative neighborhood aggregation process, which updates node embeddings based on the information aggregated from neighboring nodes. This is achieved through a message-passing framework that recursively updates node embeddings. During each iteration, nodes aggregate information from their neighbors by using a neural network-based aggregation function, and their embeddings are updated accordingly. The aggregation process can be formally described as:

\begin{equation}
{\mu_i}^{(t)} = AGGREGATE({{\mu_j}^{(t-1)} : j \in N(i)})
\end{equation}

where ${\mu_i}^{(t)}$ is the updated embedding of node $i$ at iteration $t$, $N(i)$ denotes the set of neighbors of node $i$, and AGGREGATE is the neural network-based function as in Figure 5. This process is repeated for a fixed number of iterations T, culminating in the final embeddings ${{\mu_i}^{(T)}: i \in V}$ as the area presentation.

\begin{figure}[h]
 \centering
  \includegraphics[width=5.0in]{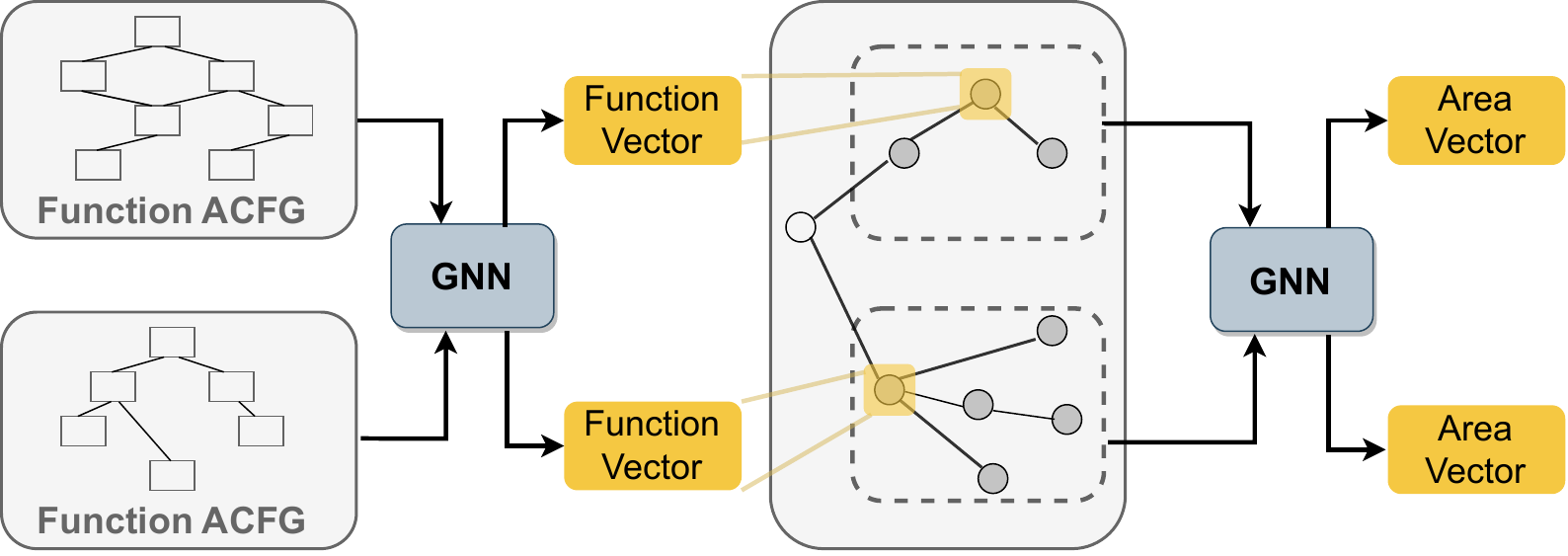}
  \caption{The architecture of Embedded-GNN. The GNN is a Structure2vec network, which is the same as in Figure 5. We use the Function Vector for anchor detection in Section 3.1 and use the Area Vector to calculate the Area structure score in Section 3.3.1.}
\end{figure}

Finally, Structure2Vec employs a supervised learning objective that maximizes the similarity between embeddings of homologous areas. In detail, we initially selected all the binary FCGs in Dataset\_OSS and sampled areas on those FCGs. We selected the nodes with more than 5 adjacent nodes on FCG as the initial nodes for each area.
For training the GNN for FCG embedding, we treated the homologous areas as positive samples and utilized the areas obtained from non-homologous FCG sampling as negative samples. 

After that, we leveraged the trained reuse area embedding model to generate vector representation for function areas after the Area Generation module. Similarly, we regard the vector cosine similarity between the target area and TPL area as their structural similarity score $S$. Two areas are considered structurally similar if $S$ is greater than a threshold value of 0.8. The threshold value is set in Section 4.

\subsubsection{Anchor Alignment} 

After calculating structural similarity, numerous non-homologous functions with similar structures and similar areas still exist, leading to false positives and false negatives in TPL detection results. Considering the importance of one-to-one matching of nodes between two areas in the graph-matching domain, we explore the potential for improved TPL detection by considering the number of matched anchor pairs. Intuitively, an area is more likely to be a reused area if it has a higher number of matched anchor pairs. We refer to the approach in previous work LibDB \cite{LibDB}, where they determine the reuse by detecting whether the matched nodes on the FCG have three common edges. However, LibDB \cite{LibDB} suffers from an \textbf{\textit{overlapping}} problem, which is discussed in detail below. In order to solve the overlapping problem and compute a more accurate number of aligned nodes between areas, we propose an Anchor Alignment algorithm to compute the similarity by computing the maximum length of a one-to-one alignment relation of matched nodes in two areas.

One of the challenges encountered is the \textbf{\textit{overlapping}} phenomenon. Because all function pairs whose similarity is larger than the threshold are seen as anchor pairs during the anchor detection phase, one target function may be associated with multiple candidate TPL functions while other target functions may only be related to one candidate node, which is called \textbf{\textit{overlapping}}. For example, in Figure 8(a), nodes A and C on the TPL FCG have three matched nodes on the Target FCG. To address this issue, a naive approach would be to enumerate all matching combinations to obtain all groups of aligned areas. However, this approach is computationally intensive and time-consuming due to the vast number of possible combinations.

To tackle this problem, we propose the Anchor Alignment algorithm, which calculates the longest list of anchor pairs without overlapping. We define the length of the longest anchor pair list as the \textbf{\textit{alignment length}}. In Figure 8, the ideal list is [(1, A), (2, B), (3, C), (4, D)] and the alignment length is 4. 

\begin{figure}
  \centering
  \subfigure[FCG with Overlapping]{
  \includegraphics[width=2.6in]{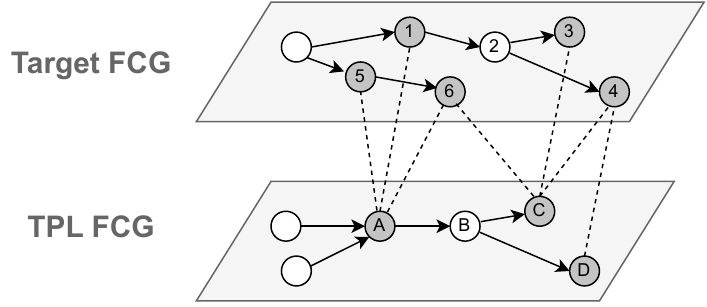}
  }
  \hspace{0.5cm}
  \subfigure[FCG after Anchor Alignment]{
  \includegraphics[width=2.6in]{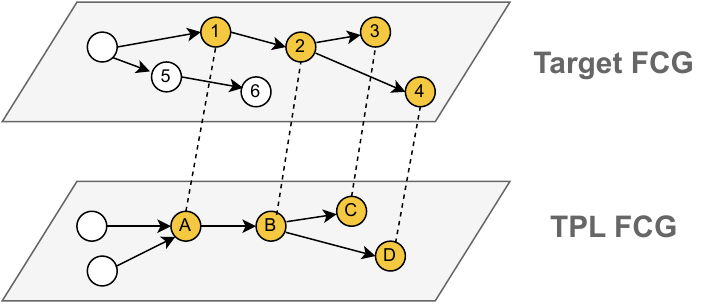}
  }
  \caption{8(a) is the target FCG (Above) and the TPL FCG (Below) before Anchor Alignment while 8(b) is after that. The gray nodes are anchors and the yellow nodes are aligned nodes. The dotted lines are the matched relations of anchors.}
  \Description{The overlapping phenomena without anchor alignment.}
  \vspace{-1.0em}
\end{figure}

We refer to the RANSAC algorithm \cite{RANSAC} in the field of Image Alignment \cite{graphalignment1, graphalignment2}, which aims to find the mapping relationships of points in two images. Nevertheless, Anchor Alignment is different from Image Alignment in that we do not calculate the Homograph Matrix \cite{matrix} (a 3*3 matrix used to rotate all points in one image to another), but the alignment length. The pseudo-code for Anchor Alignment is in Algorithm 1. The function $get\_descendants$ is to get all descendant nodes. We use the function $random.sample$ in Python to get the random anchor from nodes in the area. The function $get\_related\_tpl$ aims to get the anchor pair of target anchor $a_{tar\_child}$ with the matched TPL anchor in $A_{tpl\_desc}$ that has the most descendant nodes.

First, every candidate TPL from the Area Generation module has an anchor pair list $A$, and its members are anchor pairs that correspond to the dashed line in Figure 8. The anchor pair list in Figure 8 is [(5,A),(1,A),(6,A),(6,C),(3,C),(4,C),(4,D)]. Then, we perform the Anchor Alignment algorithm for each anchor pair in $A$. We take the anchor pair as the current generation and iteratively search for the next generation anchor pair and use it as the new current generation. As in lines 15-28 in Algorithm 1,  we randomly select an anchor pair ($a\_tar$, $a\_tpl$) in the area of the current generation as the 
\IncMargin{1em}
\begin{algorithm}[H]
        \SetAlgoLined 
	\caption{Anchor Alignment}
	\KwIn{anchor pair$ (a_{tar},a_{tpl})$, anchor pair list $A$, target FCG $g_{tar}$ and TPL FCG $g_{tpl}$}
	\KwOut{the max number of aligned anchor pairs $l_{max}$}
	\BlankLine
        
        \SetKwProg{Def}{Function}{:}{end}
        \SetKwFunction{}{}
        \Def{AnchorAlignment ($\ a_{tar},\ a_{tpl},\ A,\ g_{tar},\ g_{tpl}\ $)}{
        \SetAlgoVlined
            $l_{max}\  =\  0, \ \ n\  =\  0$\;
            \While{$True$}
            {
                $l = Alignment (\ a_{tar},\ a_{tpl},\ A,\  g_{tar},\  g_{tpl}\ )$\;
                \uIf{$l\  >\  l_{max}$}{
                   $l_{max}\  =\  l,\ \ n\  =\  0$\;
                }\Else{
                    $n\ +=\  1$\;
                }
                \If{$n\  >=\  100$}{
                    $break$\;
                }
            }
            return $l_{max}$ \;
        }
        \BlankLine
        \SetKwProg{Functionl}{Function}{:}{end}
        \Functionl{Alignment ($\ a_{tar},\ a_{tpl},\ A,\  g_{tar},\  g_{tpl}\ $)}{
            $N = 0$\;
            $A_{tar\_desc}\  =\  get\_descendants (\ a_{tar},\  g_{tar}\ )$\;
            $A_{tpl\_desc}\  =\  get\_descendants (\ a_{tpl},\  g_{tpl}\ )$\;
            $a_{tar\_child}\  =\  random.sample (\ A_{tar\_desc},\  1\ )$\;
            $a_{tpl\_child\ } =\  get\_related\_tpl (\ a_{tar\_child},\  A, \ A_{tpl\_desc}\ )$\;
            \uIf{$\ is\_not\_empty(\ a_{tpl\_child}\ )\ $}{
                   $N\  +=\  1 + Alignment (\ a_{tar\_child},\  a_{tpl\_child},\  A, \ g_{tar},\  g_{tpl}\ )$\;
            }\Else{
                return $N$\;
            }
            return $N$\;
        }

\end{algorithm}
\noindent next generation and make sure they are not overlapped. For instance, (3, C) and (4, D) are not overlapped while (3, C) and (4, C) are overlapped. Next, we record the alignment length $l$ of the new list if $l > l_{max}$ and re-select iteratively. Finally, we stop the iteration when the number of iterations $n$ reached 100 without $l_{max}$ updating.

After alignment, we obtain the longest alignment length for each area. In Figure 8, the ideal list is [ (1, A), (3, C), (4, D)]. Since node (2, B) misses in the anchor detection phase, the final longest alignment length is 3. However, LibAM is robust to misses and false positives of function matching, which does not affect the final result too much. We then set the threshold $n$ and judge to reuse when the alignment length $L$ is more than $n$. Based on the evaluation in Section 4, we set $n=3$. By employing the Anchor Alignment algorithm, we can effectively address the overlapping issue, contributing to a more accurate and efficient TPL detection process.

Finally, we consider areas with anchor function similarity, area structural similarity, and area alignment length all above the thresholds as reuse areas. We record the list of reused TPLs for the TPL detection task and the list of reused functions for the Area detection task.

\subsection{Optimization Strategies} 

Moreover, we discuss the optimization strategies for both the TPL detection and Area detection tasks. Our aim is to make LibAM lightweight and efficient while maintaining its accuracy and robustness. We incorporate various techniques, such as the Annoy vector search engine, to speed up the comparison of functions and areas, as well as employ a speedup strategy for the anchor alignment algorithm. By doing so, LibAM can efficiently handle situations where the size and number of areas are huge. These optimizations are detailed below:

For the TPL detection task, in each combination of target and TPL binaries, we iteratively select a pair of candidate areas randomly to compute their structural similarity and alignment length. Once both the structural similarity (S) and alignment length (A) of the candidate area surpass the designated thresholds, we can conclude that the target binary reuses the TPL and terminates the loop. Then, we move on to the next combination. This strategy significantly reduces the computational overhead and speeds up the TPL detection process. No matter how large the area is, once the alignment length exceeds the threshold, LibAM can return the result.

For the area detection task, we only perform area detection for the detected tuples (target binary, TPL binary) after the TPL detection task. We begin by ranking the areas generated during the Area Generation module in descending order, based on their size. When a larger area is identified as reused, any smaller areas contained within it are not repeatedly computed, which further avoids the impact of an excessive number of areas. Moreover, if an area matches with multiple areas simultaneously, we select the one with the highest alignment length and structural similarity score as the final reused area. This approach not only reduces redundancy in computation but also ensures a more accurate area detection result.

\section{Evaluation}
In this section, we elaborate on the evaluation process of our proposed approach, LibAM, and discuss its accuracy in various tasks as compared to state-of-the-art (SOTA) works. We provide a comprehensive analysis, including an investigation into the bad cases of existing works, and identify the reasons for their shortcomings. Before that, we describe our experimental setup, the dataset used, and the tools employed for disassembly and coding.

In the experiments, we aim to answer the following research questions:

\textbf{RQ1}: How does LibAM perform in the TPL detection task in the public real-world dataset compared to other works?

\textbf{RQ2}: How does LibAM perform in the Area detection task in the public real-world dataset compared to other works? 

\textbf{RQ3}: Is LibAM robust in detecting TPLs in different optimization options and architectures?

\textbf{RQ4}: How do the individual components of LibAM impact the final accuracy? 

\textbf{RQ5}: How efficient is LibAM compared to other works?

\textbf{RQ6}: How does LibAM perform in detecting TPLs in large-scale real-world firmware?

First, let us describe the experimental setup and the dataset used for evaluation. We employed a high-accuracy computing environment to ensure accurate and efficiency evaluation. The system runs on Ubuntu 22.04 and is equipped with an Intel Xeon 128-core 3.0GHz CPU, which includes hyperthreading capabilities. In addition, it has 1TB of RAM and two NVIDIA V100 32GB GPUs. This setup guarantees ample resources for the execution and comparison of different algorithms. In fact, LibAM is lightweight and can even run on machines without a GPU. For the disassembly of binaries, we used IDA Pro 6.8, a widely recognized disassembler and debugger. IDA Pro 6.8 has extensive capabilities to reverse-engineer and analyze binary files, making it an ideal choice for our evaluation process. The code for our proposed method is primarily written in Python 3.6, with the IDAPython code being in Python 2.7. This choice of languages allows us to leverage the extensive library support and ease of use that Python offers.

In this evaluation section, we assess the accuracy of our proposed approach, LibAM, by using widely-accepted metrics, namely Precision (P), Recall (R), and F1 score (F1). These metrics are suitable for determining the accuracy of our approach in detecting TPLs and reuse areas. In addition, we performed statistical analysis of the results using violin plots and CDF (Cumulative Distribution Diagram). In detail, $TP$ presents the number of reused TPLs that are correctly detected as reused while $TN$ presents the number of unused TPLs that are correctly detected as unused. Besides, $FN$ presents the number of reused TPLs that are incorrectly detected as unused while $FP$ presents the number of unused TPLs that are incorrectly detected as reused. Moreover, the equations of Precision, Recall and F1 score are as follows:

\begin{equation}
Precision = \frac{TP}{TP + FP}
\end{equation}

\begin{equation}
Recall = \frac{TP}{TP + FN}
\end{equation}

\begin{equation}
F1 = 2\cdot\frac{Precision\cdot Recall}{Precision+Recall}
\end{equation}

To maintain consistency and facilitate comparison, we preserve the vector dimension of both the function and the Function Call Graph (FCG) at the same level as in Gemini \cite{Gemini}, with a dimensionality of 64. This allows for a fair comparison between the accuracy of our approach and other Gemini-based works.

\begin{figure}[h]
  \centering
  \includegraphics[width=2.8in]{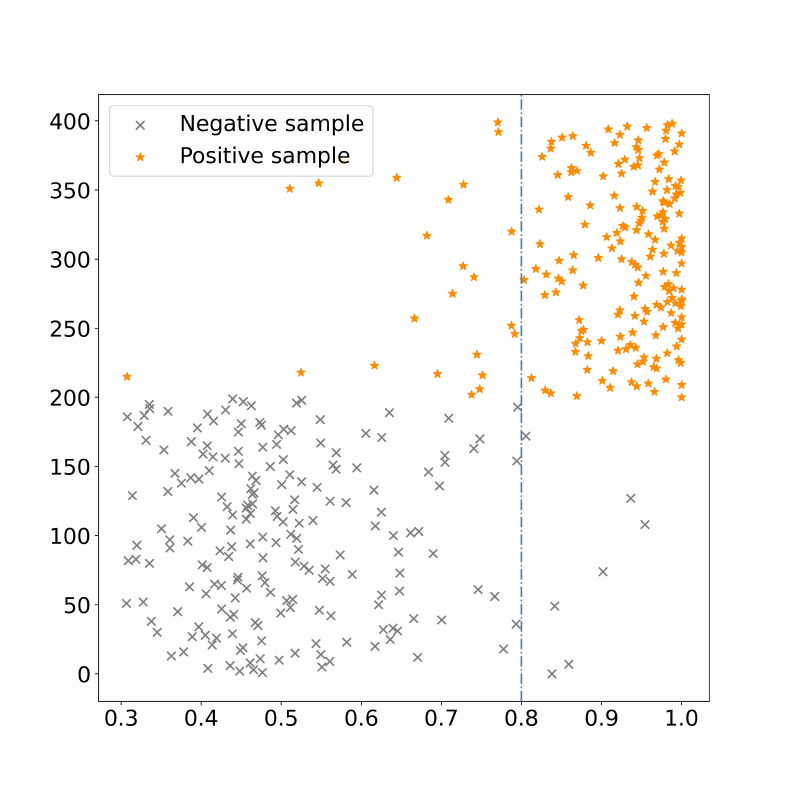}
  \includegraphics[width=2.8in]{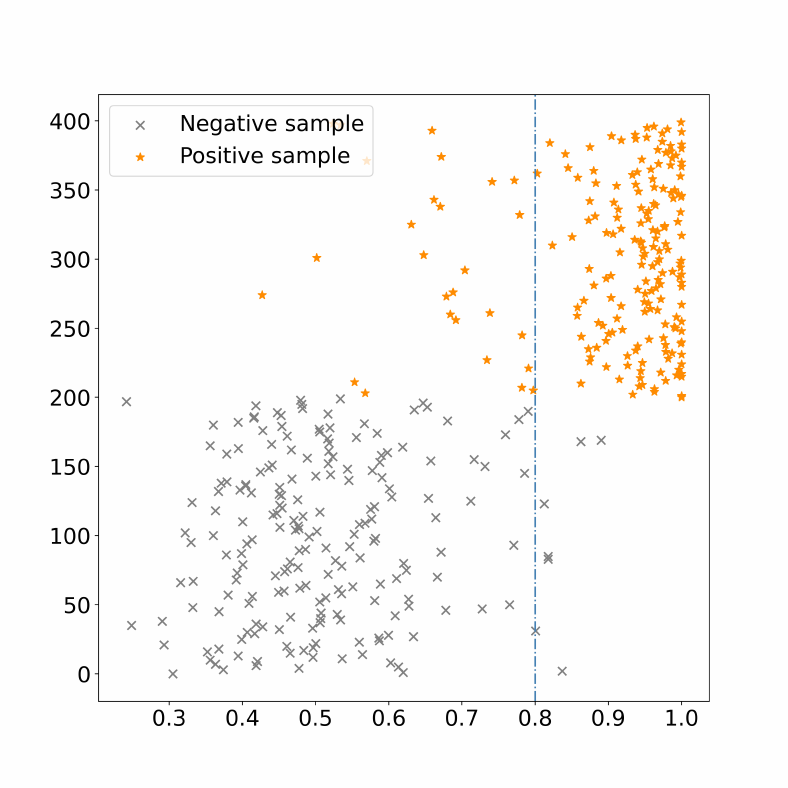}
  \caption{The similarity score of 400 random function pairs (Left) and area pairs (Right).}
\end{figure}

To identify anchor functions and assess the structural similarity of the FCG, we apply optimal threshold values derived from the validation dataset in Dataset-I, which are 0.72 and 0.8 respectively. These threshold values were determined through experimentation and analysis on the validation dataset to achieve the best balance between precision and recall.

As shown in Figure 9, we randomly select a subset consisting of 400 function pairs and FCG pairs from Dataset-I to provide a representative sample for our evaluation. The selected thresholds, 0.72 and 0.8, effectively distinguish between positive (homologous) and negative (non-homologous) samples. This indicates that the threshold values are adept at identifying homologous code while minimizing false positives and false negatives.

Furthermore, we set the alignment length threshold based on the number of common edges employed in LibDB \cite{LibDB} with a value of 3, thus facilitating a comparison of the effectiveness of the Anchor Alignment algorithm compared to the simple common edge filtering rule in LibDB \cite{LibDB}. This threshold serves as an additional criterion to ensure that the identified reusable TPLs have a sufficient level of similarity. We further evaluated the impact of different values of this threshold on LibAM in Section 4.6.

\subsection{Dataset}

In order to comprehensively evaluate the accuracy of LibAM in different environments, we leveraged one independent dataset for training our model and three additional datasets for evaluation purposes. The details of these datasets are presented in Table 1.

\textbf{Dataset\_OSS}:
To develop a rich and diverse dataset for training our function embedding model and Embedded-GNN network, we invested significant time and effort into crawling 260 commonly used open-source projects from Github \cite{github} and SourceForge \cite{sourceforge}. We manually compiled these projects into 22,100 binaries, encompassing three architectures (ARM, x86, x64) and four optimization options (O0, O1, O2, O3). This extensive dataset ensures that our model is well-trained and capable of handling various real-world scenarios. We divide it into a training set, validation set, and testing set in a ratio of 8:1:1.

\textbf{Dataset\_ISRD}:
To evaluate the ability of LibAM on existing public real-world datasets, we leverage a real-world reuse dataset used by ISRD \cite{ISRD}. The dataset contains 85 binaries from 24 popular open-source projects across various domains, compiled with default optimization options in x64, and includes 74 real partial reuses. This dataset is the only complete public TPL detection dataset as we know. While another TPL dataset used in LibDB \cite{LibDB} only contains a ground truth file without corresponding binaries. Even though these target binaries originate from the FedoraLib dataset  \cite{LibDB}, their names and version numbers in the ground truth file don't correspond to those in the FedoraLib dataset, posing difficulties in utilizing this dataset for TPL detection. (for example, \textit{vorbis} in ground truth file while there are many confused binaries like \textit{libsox\_fmt\_vorbis.so}, \textit{libvorbis.so.0.4.6}, \textit{libvorbisenc.so.2.0.9} and so on in FedoraLib.)

\begin{table}
  \centering
  \caption{dataset scale}
  \label{tab:freq1}
  \begin{tabular}{cccccc}
    \toprule
    Dataset & Binaries & Target Binaries & TPLs & Contents & Applications\\
    \midrule
    Dataset\_OSS & 22100 & - & - & cross arch and opti & software for traning\\
    Dataset\_ISRD & 85 & 17 & 85 & real-world & software for TPL detection\\
    Dataset\_ExtISRD & 289 & 204 & 85 & cross arch and opti & software for TPL detection\\
    Dataset\_FW & 12915 & 12699 & 216 & real-world & firmware for TPL detection\\
  \bottomrule
  \vspace{-2.0em}
\end{tabular}
\end{table}

\textbf{Dataset\_ExtISRD}:
Recognizing the limitations of existing TPL detection datasets, we expanded the ISRD dataset by manually compiling binaries for three architectures (arm, x86, x64) and four optimization options (O0, O1, O2, O3). This extended dataset, with 289 binaries and 477 real partial reuses, enables us to assess the accuracy of LibAM under a broader range of conditions, thus providing a more comprehensive evaluation of its capabilities.

\textbf{Dataset\_FW}:
To evaluate LibAM's ability to tackle large-scale firmware TPL detection tasks, we collected 167 firmware from 10 different vendors. Such kinds of firmware spanned various device categories, such as IP cameras, routers, and switches. We used Binwalk \cite{Binwalk} to decompress the firmware and extracted 12,699 binaries. This large-scale dataset allowed us to test the scalability and applicability of LibAM in real-world situations, further demonstrating its scalability and practicality.

By leveraging these four datasets, we were able to rigorously evaluate the accuracy of LibAM in a variety of environments, ensuring that our model is adaptable and effective in handling real-world challenges. The results from this comprehensive evaluation serve as strong evidence for the suitability of LibAM for deployment in practical TPL detection tasks.

\subsection{Compared Methods}
In this evaluation, we introduced the comparison between LibAM and various existing methods, examining their strengths and weaknesses in different scenarios. We aimed to provide a comprehensive assessment of the accuracy of both constant-based works and function similarity-based works, and to demonstrate the effectiveness of LibAM. The comparison methods are further elaborated on below:

\textbf{LibDX}: LibDX \cite{LibDX} has the advantage of being a simple and fast TPL detection approach. However, its reliance on constant features, specifically strings, might result in limited detection capabilities when dealing with binaries with few strings. Additionally, this approach may produce false positives when the logical blocks in the target binary and TPL are coincidentally similar.

\textbf{B2SFinder}: B2SFinder \cite{B2SFinder} boasts a comprehensive range of constant features, allowing it to perform a more in-depth analysis when detecting TPL reuse. However, this approach may suffer from increased computational complexity and time cost due to the increased number of features it employs. Moreover, as a constant-based work, B2SFinder still faces performance degradation from binary with fewer constant features.

\textbf{ISRD$_{Gemini}$}: ISRD \cite{ISRD} checks if more than half of the functions in the TPL match the functions in the target binary to determine reuse. However, as ISRD is limited to detecting TPLs in cross-architecture environments and has no open-source version available, we did not perform a direct comparison between LibAM and ISRD. Instead, we use Gemini \cite{Gemini} as a baseline with the strategy of ISRD, which is called $ISRD_{Gemini}$. Note that, LibAM focuses on the work after function matching and can simply replace Gemini with ISRD to detect anchors in a single environment. 

\textbf{LibDB}: LibDB \cite{LibDB} adds FCG information to perform simple filtering of function matching results. It detects if there are more than three matched functions on the FCG to determine reuse. Unlike LibDB, which performs a simple filter on isolated functions, LibAM connects the isolated functions into areas to compare the structural similarity of areas and solves the overlapping problem of LibDB with the Anchor Alignment algorithm.

By comparing LibAM with these existing works, we were able to highlight the advantages and unique features of our approach. As for ModX \cite{ModX}, since it is not open source and it focuses on program modularity and reverse program semantic understanding, TPL detection is only one of its application tasks, we did not implement it. Through this evaluation, we demonstrated that LibAM outperforms other methods in various aspects, such as the ability to detect TPLs across different environments, offering a comprehensive analysis through the Area detection task, and addressing the overlapping issue using the Anchor Alignment algorithm. This comparison demonstrates the effectiveness and superiority of LibAM in TPL detection tasks.

\subsection{Answer to RQ 1: Accuracy of LibAM in public real-world dataset}
In this evaluation, we test the accuracy of LibAM on public real-world Dataset\_ISRD. Table 2 reports the accuracy of LibAM in terms of precision, recall, and F1 score compared with other works.


The results showed that LibAM achieved the highest precision and recall, with a precision of 1.0 and a recall of 0.993. Compared to other works, LibAM's precision was 8.0\% higher than the second-ranking LibDX \cite{LibDX}, and the recall was 7.3\% higher than the second-ranking LibDB \cite{LibDB}. LibAM can accurately identify every reuse in Dataset\_ISRD, and only the \textit{liblzg} code area reused by \textit{lzbench} is too small to cause LibAM to miss it. The reuse area detection in Section 4.4 can further explain the good results of TPL detection.

To our surprise, LibDX \cite{LibDX} achieves the second-highest precision of 0.92 and a high recall of 0.809, demonstrating that continuous strings in binaries can provide weak semantic information. The precision of B2SFinder\cite{B2SFinder} that uses more types of constant features is only 0.692 in precision and 0.644 in recall. Although B2SFinder uses many types of features rather than just strings, the use of thresholds for the entire file granularity makes both precision and recall significantly lower than LibDX. Although LibDX can achieve good scores, it is still significantly lower than LibAM due to the harsh conditions for ten continuous strings. For example, LibDX\cite{LibDX} can't detect the \textit{lzbench} reuse from \textit{csc}, because even if the number of common strings is large, the number of continuous strings is only 2. 

\begin{table}
  \centering
  \caption{Accuracy on TPL detection task on Dataset\_ISRD}
  \label{tab:freq2}
  \begin{tabular}{cccccc}
    \toprule
     Model & LibDX & B2SFinder & $ISRD_{Gemini}$ & LibDB& LibAM\\
    \midrule
    P & 0.920 &  0.692 & 0.304 & 0.534 &\textbf{1.0}\\
    R & 0.809 &  0.644 & 0.518 & 0.920 & \textbf{0.993}\\
    F1 & 0.837 &  0.664 & 0.312 & 0.568 & \textbf{0.996}\\
  \bottomrule
\end{tabular}

\end{table}

We analyzed the bad cases and find that both constant-based works \cite{LibDX, B2SFinder} have limitations in real-world scenarios for two main reasons. Firstly, some reused binaries have few constant features, such as the \textit{brotli} code reused in \textit{lzbench}. Additionally, some binaries slightly modify the reused code to remove string-print instructions without changing the semantics. As in the case of \textit{minizip}, which removes the string-print instructions in the \textit{$BZ2\_BlockSort$} function of \textit{bzip2}. Consequently, both constant-based works fail to detect this reuse relation. To deal with this problem, LibAM uses function matching and area matching to detect functions that do not contain or contain few strings.

Both the precision and recall of $ISRD_{Gemini}$\cite{ISRD} are the lowest which proved that directly using function similarity result as the TPL detection result can't get a satisfactory result. Although Figure 9 demonstrates that Gemini has a high ability to distinguish between pairs of functions that are homologous and non-homologous, it is quite difficult to retrieve homologous functions in large-scale functions, which is consistent with the results of jTrans \cite{jTrans}.

To handle this problem, LibDB \cite{LibDB} leverages common edges in FCG to filter the function similarity results and get a higher recall. However, LibDB \cite{LibDB} only uses three common edges to filter functions, which is too simple to cause many false positives. Moreover, the overlapping phenomenon further aggravates the problem.

On the contrary, based on the function matching results, LibAM expands the comparison granularity to the area on FCG and further detects the similarity of areas using area structural similarity and anchor alignment length, so as to obtain good results.

\begin{tcolorbox}[colback=gray!10,
                  colframe=black,
                  arc=1mm, auto outer arc,
                  boxrule=1.5pt,
                 ]
\textbf{Answering RQ1:} LibAM effectively detects TPLs in the public real-world dataset and outperforms all existing works with a precision of 1.0 and recall of 0.993, which manifests that LibAM is powerful at detecting reused code from a large amount of non-reused code.
\end{tcolorbox}

\subsection{Answer to RQ 2: Accuracy of detecting reuse areas and the interpretable evidence for TPL detection}

In this evaluation, we employ Dataset\_ISRD to evaluate the exact area detection ability of LibAM. Three of us manually analyzed the exact reuse areas as ground truth over the course of a week. Since neither the target binaries nor the TPL binaries in the ISRD dataset have deleted function names, we can easily analyze which functions are reused and generate ground truth by function names. However, the function names in some reuse areas may be slightly modified and cannot be directly compared by string, so we manually filter them one by one to determine the exact reuse areas. 

\begin{table}
  \centering
  \caption{Accuracy on Area detection task on Dataset\_ISRD}
  \label{tab:freq3}
  \begin{tabular}{cccccc}
    \toprule
     Model & LibDX & B2SFinder & $ISRD_{Gemini}$ & LibDB& LibAM\\
    \midrule
    P & 0.779 &  0.519 & 0.250 & 0.613 &\textbf{0.985}\\
    R & 0.291 &  0.372 & 0.573 & 0.719 & \textbf{0.847}\\
    F1 & 0.379 &  0.393 & 0.311 & 0.619 & \textbf{0.910}\\
  \bottomrule
\end{tabular}

\end{table}

As shown in Table 3, LibAM can accurately detect reuse areas by filtering at the function level, area structure level, and area node level of 0.985 in precision and 0.847 in recall. In order to avoid false positives, we set the areas with an alignment length of more than 3 to be recognized as reuse areas, which makes some small areas easy to be missed, thus leading to the fact that recall is not as high as precision. However, the results show that LibAM is able to detect accurate reuse areas to support the interpretable TPL detection results.

The previous works only performed TPL detection without Area detection, so they cannot be used directly for area detection, we simply modified existing work to support area detection and compare them with LibAM. Specifically, we treat the functions that use the matched constant features in LibDX and B2SFinder as reuse areas. For $ISRD_{Gemini}$, we just take matched functions as the reuse area. For LibDB, we use the functions that satisfy the three common edges as the reuse area.

LibDX \cite{LibDX} still has the second-highest precision of 0.779, but its recall is only 0.291. Our analysis of bad cases shows that a large number of matched strings are not called by functions in LibDX, which makes it difficult to determine which piece of code is being reused. 

The same problem occurs in B2SFinder \cite{B2SFinder}, which uses a wider variety of features and has a significantly higher recall than LibDX, but still only 0.372. This demonstrates that while Constant-based works are convenient and effective, it is difficult to identify which code areas are actually reused, and these matched features are sometimes difficult to use as interpretable evidence that TPL is actually reused.

Function similarity-based works have a significantly higher recall due to the comparison of all functions. Even $ISRD_{Gemini}$, which directly uses the matched functions as the reuse area, has a recall of 0.573. After filtering with 3 common edges, LibDB obtains a precision of 0.613 and a recall of 0.719. However, there is still a big gap between them and LibAM.


\begin{tcolorbox}[colback=gray!10,
                  colframe=black,
                  arc=1mm, auto outer arc,
                  boxrule=1.5pt,
                 ]
\textbf{Answering to RQ2:} Compared to previous methods, which failed in the exact reuse area detection task, LibAM has demonstrated the feasibility of doing so with 0.985 precision and 0.847 recall. This provides interpretability of the TPL detection results and is beneficial for downstream tasks.
\end{tcolorbox}

\subsection{Answer to RQ 3: Accuracy of LibAM in different architectures and optimization options}

In this evaluation, we aimed to assess the robustness of the works under extreme conditions by leveraging Dataset\_ExtISRD. We designed a series of experiments to test the accuracy of the works when faced with different compilation option combinations and architectural variations. Our goal was to evaluate the adaptability of the works in handling various real-world scenarios, ensuring that they are applicable across a wide range of situations.

\begin{table*}[t]
\centering
\caption{TPL detection accuracy of different optimization options in x64-x64.}
\label{table1}
\begin{tabular}{c|ccc|ccc|ccc|ccc|ccc}
\toprule[2pt]
\multicolumn{1}{c|}{\multirow{2}{*}{Model}} & \multicolumn{3}{c|}{O0-default}&\multicolumn{3}{c|}{O1-default}&\multicolumn{3}{c|}{O2-default}&\multicolumn{3}{c|}{O3-default}&\multicolumn{3}{c}{Average}\\
\cline{2-16}

\multicolumn{1}{c|}{} & P & R & F1 & P & R & F1 & P & R & F1 & P & R & F1 & P & R & F1\\
\hline


\multirow{1}{*}{LibDX} &0.55	&0.44	&0.43	&0.80	&0.61	&0.65 &0.74	&0.61	&0.63 &0.74	&0.61	&0.63	&0.71	&0.56 &0.58\\

\multirow{1}{*}{B2SFinder} &0.60	&0.39	&0.45	&0.60	&0.39	&0.45 &0.58	&0.39	&0.45 &0.58	&0.39	&0.45 &0.59	&0.39	&0.45\\

\multirow{1}{*}{$ISRD_{Gemini}$} &0.13	&0.34	&0.19	&0.31	&0.36	&0.25 &0.33	&0.36	&0.28 &0.22 &0.35	&0.27 &0.25	&0.35	&0.25\\

\multirow{1}{*}{LibDB} &0.38	&0.47	&0.32	&0.44	&0.83	&0.50 &0.47	&0.85	&0.54 &0.45	&0.81	&0.51	&0.44	&0.74 &0.46\\

\multirow{1}{*}{$LibAM_{-align}$} &0.16	&1.0	&0.23	&0.17	&1.0	&0.24 &0.19	&1.0	&0.26 &0.19	&1.0	&0.28 &0.18	&1.0	&0.25\\

\multirow{1}{*}{$LibAM_{-gnn}$} &0.47	&0.86	&0.58	&0.52	&0.96	&0.62 &0.54	&0.98	&0.63 &0.58	&0.96	&0.67 &0.53	&0.94	&0.62\\

\multirow{1}{*}{LibAM}	&\textbf{0.90}	&\textbf{0.86}	&\textbf{0.88}	&\textbf{0.94}	&\textbf{0.97}	&\textbf{0.95}	&\textbf{0.94}	&\textbf{0.97}	&\textbf{0.95}	&\textbf{0.93}	&\textbf{0.96}	&\textbf{0.94}	&\textbf{0.93}	&\textbf{0.94}	&\textbf{0.93}\\

\bottomrule[2pt]
\end{tabular}
\label{table_MAP}
\end{table*}

\begin{table*}[t]
\centering
\caption{TPL detection accuracy of different architectures with O2-default options and $Mix$.}
\label{table2}
\begin{tabular}{c|ccc|ccc|ccc|ccc|ccc}
\toprule[2pt]
\multicolumn{1}{c|}{\multirow{2}{*}{Model}} & \multicolumn{3}{c|}{x86-x64}&\multicolumn{3}{c|}{x64-x64}&\multicolumn{3}{c|}{arm-x64}&\multicolumn{3}{c|}{Average}&\multicolumn{3}{c}{Mix}\\
\cline{2-16}

\multicolumn{1}{c|}{} & P & R & F1 & P & R & F1 & P & R & F1 & P & R & F1 & P & R & F1\\
\hline


\multirow{1}{*}{LibDX} &0.74	&0.61	&0.63	&0.74	&0.61	&0.63 &0.55	&0.44	&0.44 &0.68	&0.55	&0.57 &0.66	&0.52	&0.53\\

\multirow{1}{*}{B2SFinder} &0.60	&0.39	&0.45	&0.58	&0.39	&0.45 &0.60	&0.38	&0.44 &0.59	&0.39	&0.44  &0.60	&0.39	&0.44\\

\multirow{1}{*}{$ISRD_{Gemini}$} &0.33	&0.36	&0.26	&0.33	&0.36	&0.28 &0.26	&0.33	&0.28 &0.31	&0.35	&0.28 &0.24	&0.34	&0.25\\

\multirow{1}{*}{LibDB} &0.43	&0.78	&0.48	&0.47	&0.85	&0.54 &0.43	&0.63	&0.44 &0.44	&0.76	&0.49 &0.40	&0.64	&0.41\\

\multirow{1}{*}{$LibAM_{-align}$} &0.18	&1.0	&0.26	&0.19	&1.0	&0.26 &0.16	&0.99	&0.23 &0.18	&1.0	&0.25 &0.17	&1.0	&0.24\\

\multirow{1}{*}{$LibAM_{-gnn}$} &0.50	&0.98	&0.62	&0.54	&0.98	&0.63 &0.66	&0.94	&0.74 &0.56	&0.97	&0.67 &0.53	&0.91	&0.63\\

\multirow{1}{*}{LibAM} &\textbf{0.97}	&\textbf{0.97}	&\textbf{0.97}	&\textbf{0.94}	&\textbf{0.97}	&\textbf{0.95} &\textbf{0.97}	&\textbf{0.93}	&\textbf{0.94} &\textbf{0.96}	&\textbf{0.96}	&\textbf{0.95} &\textbf{0.90}	&\textbf{0.88}	&\textbf{0.88}\\

\bottomrule[2pt]
\end{tabular}
\label{table_MAP2}
\end{table*}

Table 4 presents every compilation option combination for x64-x64, where both target and TPL binaries are in x64. This setup allows us to evaluate the accuracy of the works when dealing with binaries compiled using various optimization levels and options. By examining how the works handle these diverse combinations, we can better understand their ability to cope with complex and challenging scenarios.

In Table 5, we showcase every architecture combination with O2-default options, where target binaries are compiled with O2 optimization while TPL binaries are compiled using default options. This experiment is designed to test the works' robustness when faced with discrepancies between target and TPL binaries in terms of architectures. This is particularly relevant in real-world IoT firmware.


The $Mix$ entry in Table 5 represents a combination of all 12 optimization options and architecture variations. In this experiment, we aimed to evaluate the works' accuracy under a more complex and diverse set of conditions, simulating the challenges they may encounter in real-world situations. By testing the works' adaptability to this wide range of scenarios, we can gain insights into their overall robustness and resilience.

It is important to note that the default compilation option in Dataset\_ExtISRD is a mix of O2 and O3 optimization levels, rather than a single optimization level. This choice reflects the reality of software development, where binaries are often compiled using a combination of optimization levels to balance accuracy and code size. This mixed optimization setting further enhances the complexity and diversity of the dataset, providing a more challenging testbed for the works under evaluation.


LibAM achieved a precision of 0.93 and recall of 0.94 on average for different optimization options. In different architectures, LibAM achieved a precision of 0.96 and recall of 0.96. LibAM maintains a high score across architectures. The cross-optimization option has a greater impact on LibAM, but LibAM still maintains good results even with the O0 option. These results indicate that LibAM consistently achieves high accuracy in all environments.

Due to the natural robustness of constant features under different compilation options, the results of B2SFinder are very stable. In contrast, the accuracy of LibDX \cite{LibDX} is significantly degraded due to the changes in the order of strings in different environments. The results show that the two Constant-based works are more stable across optimization options and architectures, but their scores are not ideal.

The results of both two function similarity-based works vary unstably from one compilation environment to another. LibDB has a recall rate of up to 0.85, while in the lowest case, it is only 0.47. At the limit of O0, $ISRD_{Gemini}$ only gets a precision of 0.13. This is because the homologous functions obtained by compiling in different environments change a lot, resulting in poor function similarity matching results. This indicates that the stability of TPL detection using isolated function matching is poor. In contrast, LibAM greatly reduces the impact of different compilation environments on function matching results by connecting isolated functions to areas and comparing area similarity, thus filtering out many mistakes.

\begin{table}
  \centering
  \caption{Accuracy on Area detection task on Dataset\_ExtISRD}
  \label{tab:freq3}
  \begin{tabular}{cccccc}
    \toprule
     Model & LibDX & B2SFinder & $ISRD_{Gemini}$ & LibDB& LibAM\\
    \midrule
    P & 0.478 &  0.402 & 0.131 & 0.396 &\textbf{0.941}\\
    R & 0.187 &  0.262 & 0.380 & 0.101 & \textbf{0.462}\\
    F1 & 0.269 &  0.317 & 0.195 & 0.161 & \textbf{0.620}\\
  \bottomrule
\end{tabular}

\end{table}

In Table 6, although LibAM still achieves high scores for the TPL detection task on datasets with cross-architecture and optimization options, the results for area detection drop significantly. Due to strict conditional filtering, LibAM still achieves a precision of 0.941 on the area detection task, but the recall is only 0.462. Nevertheless, LibAM still outperforms all other methods, which have unsatisfactory results on this challenging task. The reason for the significant decrease in recall is that the FCG changes across optimization options and architectures, and the Area Detection task is very demanding, so LibAM ensures high precision through strict filtering, but at the cost of some recall.

\begin{figure*}[htbp]
\begin{center}
\subfigure[Precision in different architectures]{
\includegraphics[width=2.8in]{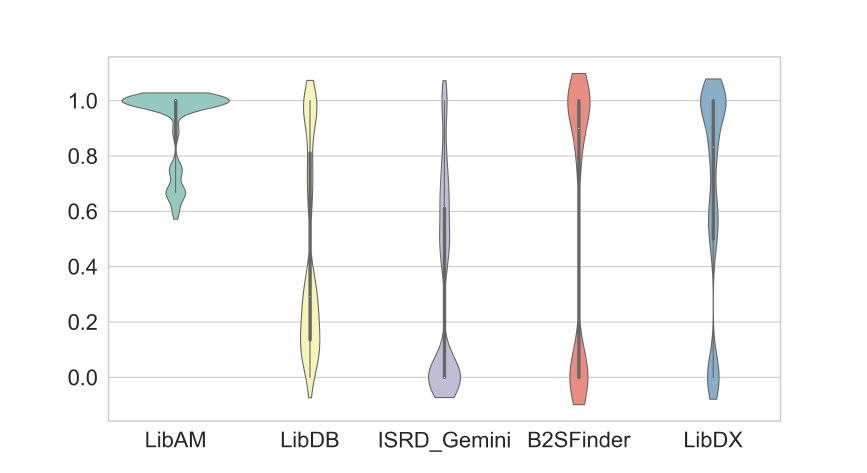}
}
\subfigure[Recall in different optimizations]{
\includegraphics[width=2.8in]{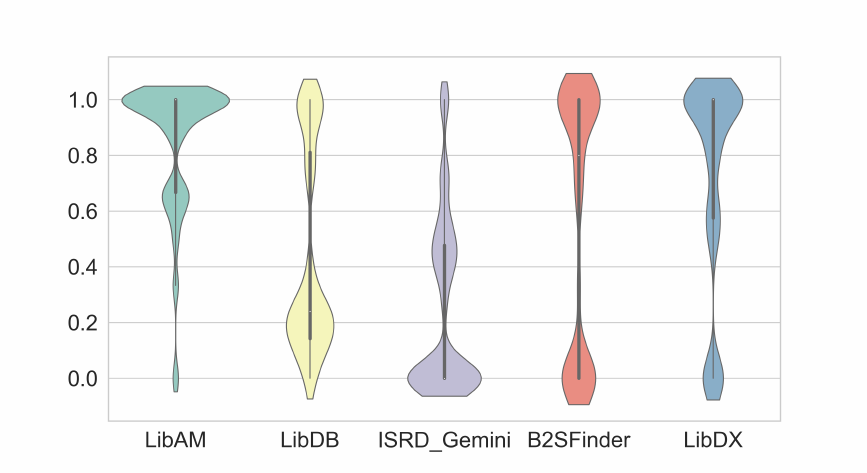}
}
\vspace{-0.8em}
\subfigure[Precision in different architectures]{
\includegraphics[width=2.8in]{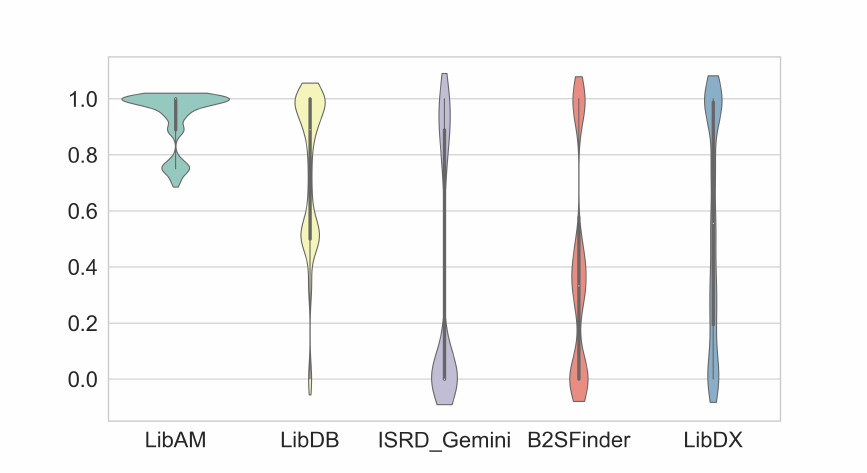}
}
\subfigure[Recall in different optimizations]{
\includegraphics[width=2.8in]{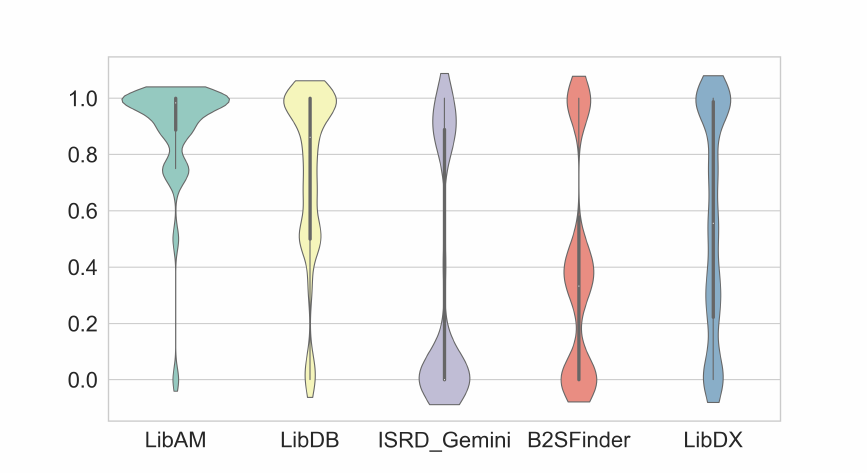}
}
\vspace{-0.8em}
\subfigure[F1 value in different architectures]{
\includegraphics[width=2.8in]{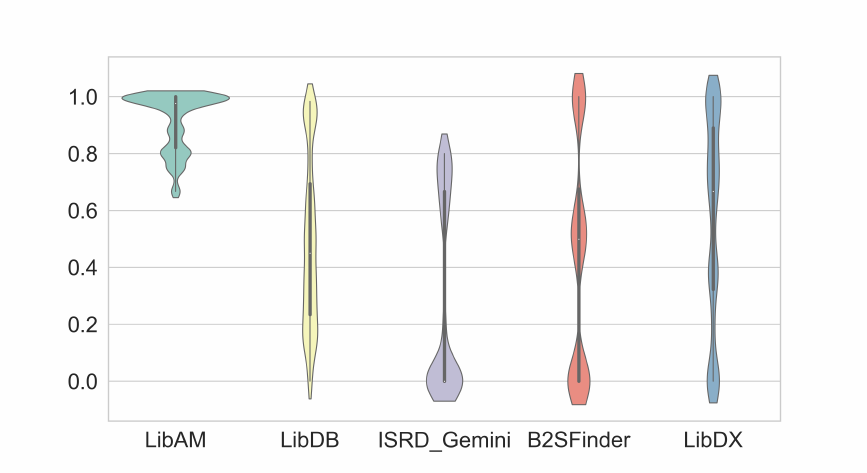}
}
\subfigure[F1 value in different optimizations]{
\includegraphics[width=2.8in]{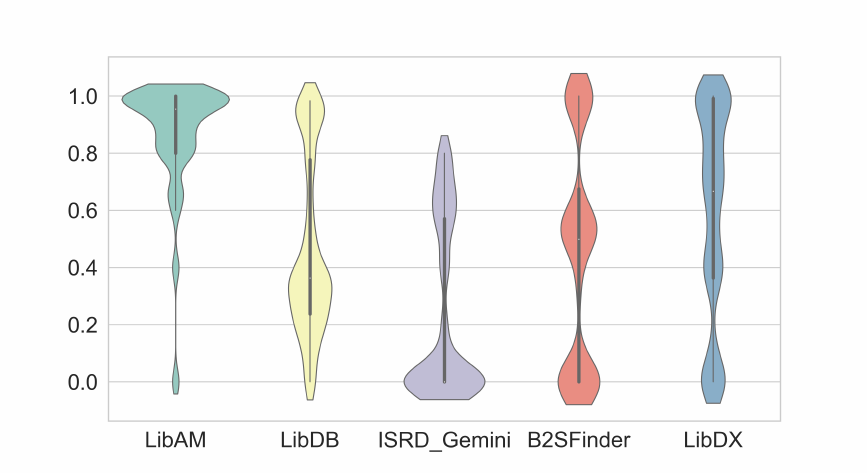}
}
\caption{Violin plots for each approach in Dataset\_ExtISRD. We have shown the distribution of Precision, Recall, and F1 value under different architectures and optimizations, respectively.}
\end{center}
\end{figure*}

In fact, detecting reused areas across optimization options and architectures is a serious challenge, and this is the first evaluation work. Therefore, the precision and recall of existing methods are less than satisfactory. Despite this challenging task, LibAM still clearly outperforms other existing methods, and the high precision rate increases the usability of LibAM. Future proposals of better function similarity matching tools may further improve the precision. Besides, new area matching methods are expected to be further proposed in the future to improve the low recall deficiency.


Figure 10 presents violin plots illustrating the distribution of precision, recall, and F1 value across various architectures and optimization options. Notably, the results of LibAM are predominantly clustered around 1.0, demonstrating a significantly superior and more concentrated accuracy compared to other methods. The results of LibDB \cite{LibDB} are mainly concentrated around 1.0 and 0.2, attributable to its high recognition accuracy for a limited number of samples that have fewer reuse relationships, while the accuracy for other samples is generally low. A substantial number of samples in $ISRD_{Gemini}$ \cite{ISRD} close proximity to 0, as relationships with a reuse proportion less than half are frequently undetected by $ISRD_{Gemini}$ \cite{ISRD}. The distribution patterns of the two Constant-based works \cite{B2SFinder, LibDX} are similar, exhibiting a polarized distribution between 1.0 and 0. This is because they have good detection accuracy for samples with rich strings, but poor detection ability for samples with fewer strings.


\begin{tcolorbox}[colback=gray!10,
                  colframe=black,
                  arc=1mm, auto outer arc,
                  boxrule=1.5pt,
                 ]
\textbf{Answering to RQ3:} LibAM is robust under different architectures and different optimization options by connecting isolated functions into areas on FCG. Optimization options have a severer impact on LibAM than architecture.
\end{tcolorbox}

\subsection{Answer to RQ 4: Impact of each part in LibAM}

In order to evaluate the impact of the Anchor Alignment algorithm and the Embedded-GNN algorithm on the accuracy of our proposed LibAM framework, we conducted two separate ablation experiments as in Table 4 and Table 5. The first variant, $LibAM_{-align}$, is a version of LibAM without the Anchor Alignment algorithm, while the second variant, $LibAM_{-gnn}$, is a version of LibAM without the Embedded-GNN algorithm. By analyzing the results of these experiments, we aim to shed light on the individual contributions of these two algorithms to the overall accuracy of the LibAM framework. In addition, we conducted sensitivity experiments on the selection of threshold $n$ for the Anchor Alignment algorithm and tested the trend of precision and recall with changes in the threshold.

In the case of $LibAM_{-align}$, after obtaining anchors via function matching, the framework solely relies on the GNN algorithm to compare the structural similarity of the areas in question. An area is deemed as a reused area if its structural similarity score surpasses a predetermined threshold value. While this approach attains a perfect recall score of 1.0 across all tested scenarios, the precision remains consistently below 0.2. This is primarily due to the presence of numerous similar Control Flow Graphs (CFGs) and Function Call Graphs (FCGs) within non-homologous functions, which leads to a high rate of false positives. These results emphasize the vital role of the Anchor Alignment algorithm in further filtering out false positives and improving the precision of the LibAM framework.

On the other hand, $LibAM_{-gnn}$ disregards the structural similarity of areas altogether. Instead, after obtaining anchors, it uses only the anchor alignment length to determine whether an area has been reused. This approach bears a resemblance to the LibDB framework but with a few key differences. Specifically, $LibAM_{-gnn}$ addresses the overlapping phenomenon present in LibDB by employing the Anchor Alignment algorithm. This results in a significant improvement in both precision and recall rates, owing to the strict one-to-one alignment relationship between anchors and the absence of proportional limitation used in LibDB. However, despite these improvements, a substantial accuracy gap remains between $LibAM_{-gnn}$ and the full LibAM framework. This indicates that considering the structural similarity of areas is instrumental in filtering out false positives and enhancing the overall accuracy of the LibAM framework.

In conclusion, our ablation experiments demonstrate the considerable contributions of both the Anchor Alignment algorithm and the Embedded-GNN algorithm to the effectiveness of the LibAM framework. The Anchor Alignment algorithm is crucial for filtering out false positives and improving precision, while the Embedded-GNN algorithm plays a significant role in further refining the identification of reuse areas. By incorporating these two algorithms, the LibAM framework achieves more robust and accurate in detecting TPLs and identifying exact reuse areas.

\begin{table}
  \centering
  \caption{The impact of threshold $n$ selection on TPL detection results in Anchor Alignment algorithm}
  \label{tab:freq4}
  \begin{tabular}{c|c|ccccc}
    \toprule
    Dataset & Metrics & n=1 & n=2 & n=3 & n=4 & n=5  \\
    \midrule
    & P & 0.604 & 0.794 & 1.0 & 1.0 & 1.0 \\
    Dataset\_ISRD& R & 0.993 & 0.993 & 0.993 & 0.947 & 0.804 \\
    
    & F1 & 0.708 & 0.856 & 0.996 & 0.971 & 0.883 \\
    \midrule
    & P & 623 & 0.671 & 0.863 & 0.811 & 0.599 \\
    Dataset\_ExtISRD& R & 0.905 & 0.896 & 0.848 & 0.452 & 0.253 \\
    & F1 & 0.704 & 0.734 & 0.842 & 0.533 & 0.322\\
  \bottomrule
\end{tabular}
\end{table}

We further investigated the impact of varying threshold values $n$ within the Anchor Alignment algorithms on LibAM's accuracy in TPL detection tasks, utilizing both Dataset\_ISRD and Dataset\_ExtISRD. For this experiment, in Table 7, n values ranged from 1 to 5. 

For Dataset\_ISRD, the results demonstrate that precision increases progressively with the rise in n, attaining its maximum at n=3. This improvement can be attributed to the algorithm's enhanced ability to differentiate between true and false reuse relationships as the threshold becomes more stringent. Conversely, recall diminishes gradually as n increases, with a substantial decline observed at n=4. This decline is likely due to the higher threshold inadvertently excluding some relevant reuse relationships. Nevertheless, LibAM demonstrates satisfactory accuracy on Dataset\_ISRD for both n=3 and n=4, achieving a balance between precision and recall.

In the case of Dataset\_ExtISRD, which comprises samples with more intricate reuse patterns, precision consistently increases with the growth of n for values less than 4. However, when n is greater than or equal to 4, the average precision begins to decline, because some samples remain undetected, which causes a precision of 0. This phenomenon suggests that a higher threshold might impede the algorithm's sensitivity to subtle reuse patterns. On the other hand, the recall still significantly decreases with the increase of n, indicating that larger thresholds are more likely to miss out on smaller reuse areas. This evaluation indicates the algorithm's capacity to accurately discern genuine reuse relationships even in more complex scenarios.

In summary, our experiments reveal that for both datasets, the optimal results are achieved when n is set to 3. This value represents a balance between precision and recall, allowing LibAM to effectively identify true reuse relationships while minimizing false positives and false negatives. These findings have implications for the tuning of the Anchor Alignment algorithm in TPL detection tasks, shedding light on the importance of selecting appropriate threshold values to optimize accuracy.



\begin{tcolorbox}[colback=gray!10,
                  colframe=black,
                  arc=1mm, auto outer arc,
                  boxrule=1.5pt,
                 ]
\textbf{Answering to RQ4:} Each component of LibAM plays a crucial role in the final results. The absence of any of these components can lead to a significant decrease in precision and recall. LibAM can get great robustness and accuracy by filtering at the function level, area structure level, and area node level.
\end{tcolorbox}

\subsection{Answer to RQ 5: Efficient of LibAM}
In this evaluation, we assessed the time cost for detecting Dataset\_ISRD and Dataset\_ExtISRD as presented in Table 8. We conducted a comparative analysis of each step in the approaches and calculated the time required for the TPL detection task and the Area detection task separately to analyze the efficiency of LibAM in relation to existing works. \textit{Feature extraction} represents the Feature Extraction phase, while \textit{Embedding} represents the function or area embedding phase. \textit{TPL detection} refers to the time cost from the completion of the Embedding phase until the TPL detection task is finished, and \textit{Area detection} refers to the time cost for the Area detection task. The full-time costs for the TPL detection task and the Area detection task are denoted as $All_{TPL}$ and $All_{area}$, respectively.

\begin{table}
  \caption{Efficiency evaluation}
  \label{tab:freq5}
  \begin{tabular}{c|cccccc}
    \toprule
    Model & Feature extraction & Embedding & TPL detection & Area detection & $All_{TPL}$ & $All_{area}$ \\
    \midrule
    LibDX & 7.5s & - & 7.5s & - & 15.1s & 15.1s\\
    B2SFinder & 7.5s & - & 70.4s & - & 77.9s & 77.9s\\
    $ISRD_{Gemini}$ & 38.8s & 1.8s & 1.1s & - & 41.8s & 41.8s\\
    LibDB & 42.5s & 1.8s & 28.6s & - & 72.8s & 72.8s\\
    LibAM & 42.5s & 3.4s & 8.3s & 109.9s & 52.7s & 162.6s\\
  \bottomrule
\end{tabular}
\end{table}

For constant-based works, the most significant time-consuming aspect is the constant comparison phase. The feature extraction phase in constant-based works, which focuses on extracting string or other constant features, is much faster than in function similarity-based works. LibDX \cite{LibDX}, which employs only string features accelerated by a backward indexing algorithm, is the fastest of all methods. On average, LibDX takes only 15.1 seconds to complete the detection of the target binary. B2SFinder \cite{B2SFinder}, on the other hand, takes a longer time to match features due to the inability to accelerate the comparison of if/switch features using inverted indexes or prefix trees like strings or arrays. B2SFinder takes 77.9 seconds to detect a target binary, making it the most time-consuming of all TPL detection methods.

For function similarity-based works, the most time-consuming process is feature extraction via IDA Pro. In $ISRD_{Gemini}$, more than 90\% of the time is spent on extracting function features. In LibDB, the TPL detection phase is also time-consuming. However, due to the acceleration strategy detailed in Section 3, LibAM is faster than both LibDB and B2SFinder in the TPL detection task, with 80\% of the time allocated to feature extraction. While LibAM is more time-consuming in the area detection task, it is the only method capable of accurately identifying reuse areas.

Furthermore, since only the detected reused relations, rather than all TPLs, are needed to further identify the reuse areas, the time consumption of the area detection phase is manageable. Moreover, as LibAM uses the top 200 TPLs and top 100 function limits, which are the same as LibDB, the detection time cost does not increase with the size of the TPL database. This aspect contributes to the overall efficiency of the LibAM framework when compared to existing works.

\begin{tcolorbox}[colback=gray!10,
                  colframe=black,
                  arc=1mm, auto outer arc,
                  boxrule=1.5pt,
                 ]
\textbf{Answering to RQ5:} 
LibAM can get high scores in a short time, which is efficient and scalable. With the optimization strategy, LibAM can even work faster than some existing works for the TPL detection task. The time consumption of the Area detection task is also in an acceptable range.
\end{tcolorbox}

\subsection{Answer to RQ 6: Accuracy of detecting large-scale reuse relation}

In this evaluation, we evaluated the ability of LibAM to detect large-scale real-world reuse relations and validated the possibilities of reuse area detection in associating vulnerabilities. Furthermore, we analyzed the detection results and made several interesting findings.

\begin{table}
  \centering
  \caption{Large-scale IoT firmware evaluation}
  \label{tab:freq6}
  \begin{tabular}{c|ccccc}
    \toprule
    Brand & Firmware & TPL & Average TPL & Vul & Average vul \\
    \midrule
    Xiaomi & 24 & 1226 & 51 & 1950 & 81 \\
    Huawei & 14 & 458 & 33 & 1893 & 135 \\
    ASUS & 25 & 808 & 32 & 2099 & 84 \\
    Dell & 25 & 192 & 7 & 1572 & 63\\
    Linksys & 12 & 141 & 12 & 1433 & 119 \\
    Dahua & 19 & 210 & 11 & 1440 & 76\\
    xiongmai & 8 & 104 & 13 & 1268 & 159\\
    Hikvision & 23 & 128 & 6 & 1485 & 65 \\
    D-Link & 9 & 53 & 6 & 1462 & 162\\
    Tplink & 5 & 33 & 7 & 261 & 52\\
  \bottomrule
\end{tabular}
\end{table}

For target binaries, we selected 10 vendors and collected 30 firmware for each vendor. Then, we use Binwalk \cite{Binwalk} to extract the file system and binaries in firmware. There are 164 firmware extracted successfully, from which we extracted 12699 binaries as in Table 9.

For TPL binaries, we first collect the widely used TPLs from Conan \cite{conan} and Vcpkg \cite{vcpkg}. Then, we compile some well-known projects from Github \cite{github} and SourceForge \cite{sourceforge} by ourselves. Besides, we also extract known TPLs in some Linux-based operating systems. Finally, we selected those with public vulnerability information and obtained 216 TPL binaries from these candidate TPLs.

LibAM detected 3353 TPLs as shown in Table 9. This entire process was completed in 30 hours. We counted the TPLs with the widest impact range, as shown in Figure 11. 11(a) shows the top 10 TPLs that are reused in the highest number of firmware and \textit{busybox} is the most reused TPL. 11(b) and 11(c) demonstrate the top 10 TPLs with the most influential vendors and binaries and they are slightly different from the TPLs in 11(a). 

In addition to detecting TPLs, we want to further associate vulnerabilities that may be introduced by TPL reuse, which is one of the downstream tasks. Firstly, we conducted a web crawler to collect the public vulnerability data of detected TPLs from CVE \cite{CVE} and NVD \cite{NVD} cites. Then, we utilized an existing technique \cite{OSSPolice} to extract all strings from the TPLs of vulnerability-related versions, enabling us to identify the specific version of the detected TPLs through string matching. After that, we associated 2519 CVEs and generated 14863 potential vulnerabilities for 167 firmware. In Figure 11(c), we listed the top 10 vulnerability types (CWE) that have the highest number of CVEs.



\begin{figure*}[htbp]
\begin{center}
\subfigure[No. of Firmware]{
\includegraphics[width=2.4in]{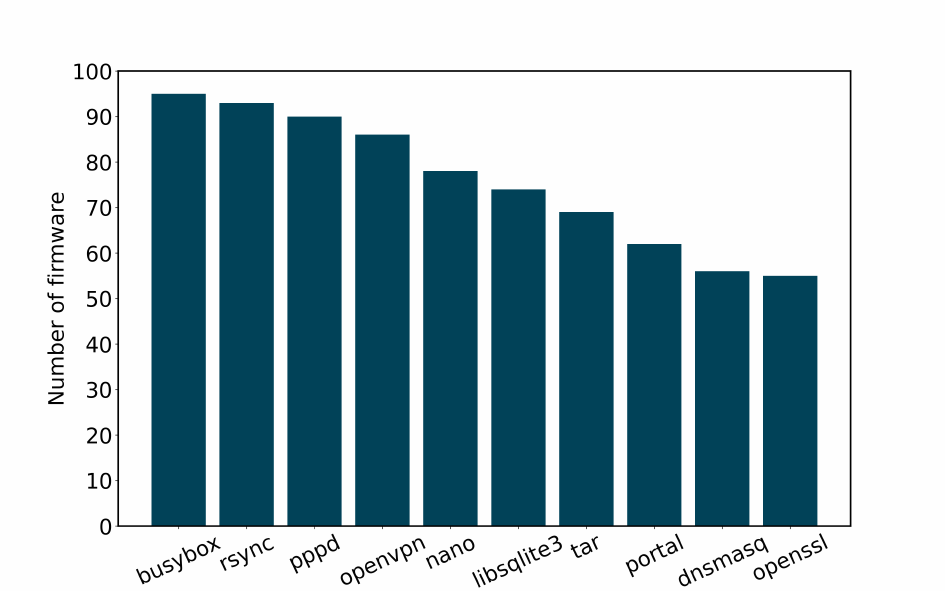}
}
\subfigure[No. of Vendors]{
\includegraphics[width=2.4in]{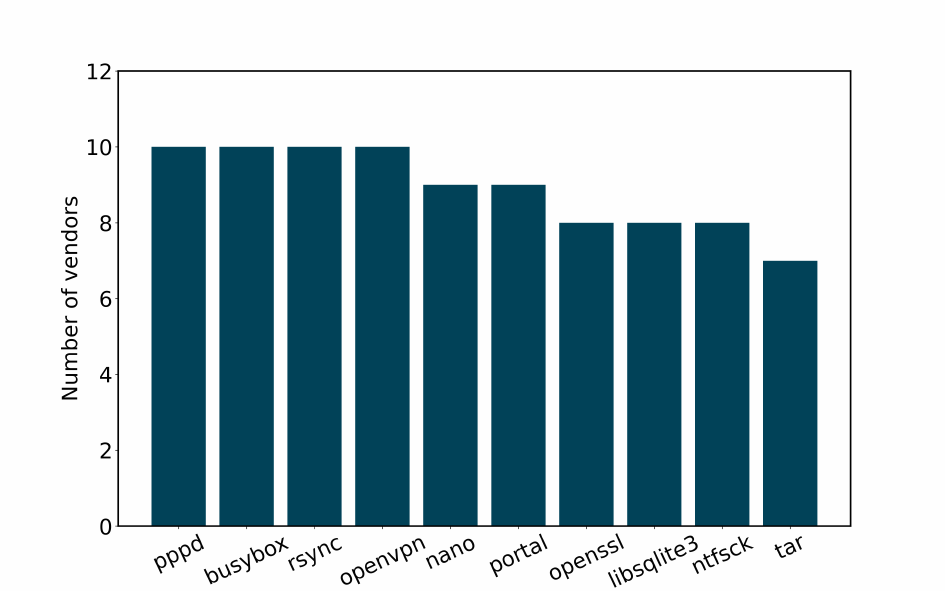}
}
\subfigure[No. of Binaries]{
\includegraphics[width=2.4in]{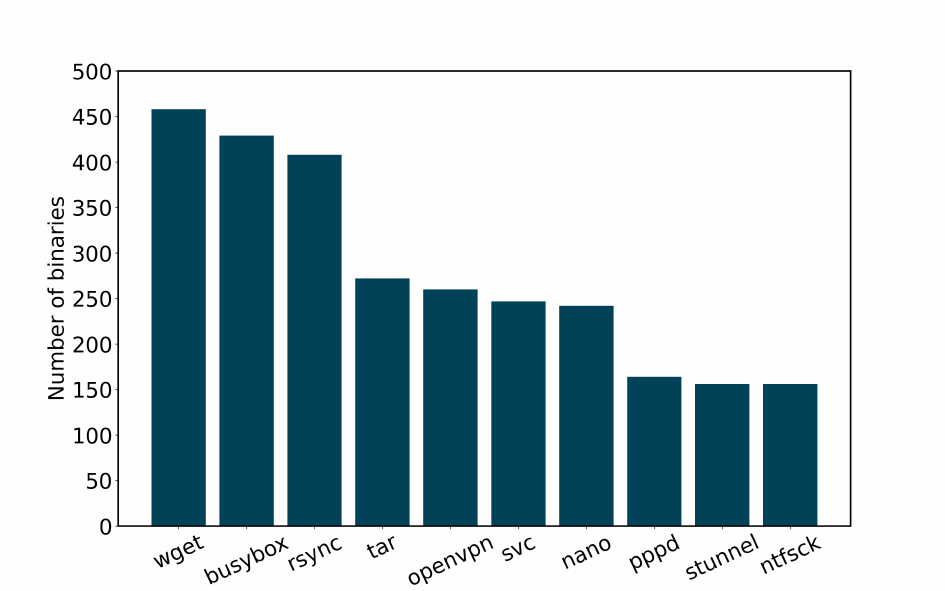}
}
\subfigure[No. of CVEs]{
\includegraphics[width=2.4in]{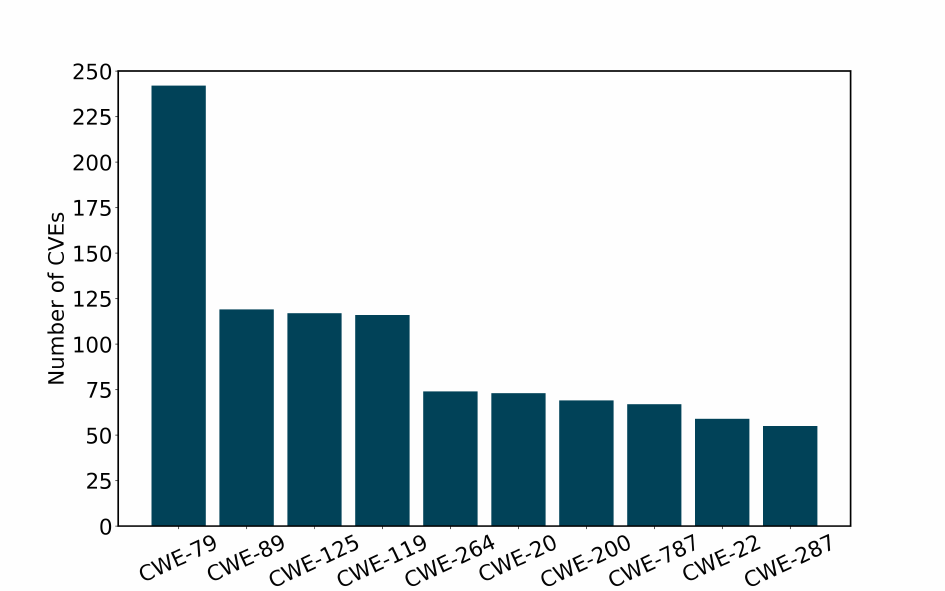}
}
\caption{The result of IoT firmware analyzed by LibAM. 11(a) is the number of firmware by the Top 10 TPLs, 11(b) is the number of vendors by the Top 10 TPLs, 11(c) is the number of binaries by the Top 10 TPLs, and 11(d) is the number of CVEs by the Top 10 CWEs.}
\end{center}
\end{figure*}

Note that many software vulnerabilities affect only partial versions, and we filter non-vulnerable versions by combining existing version identification works \cite{OSSPolice}. In addition, lots of TPLs are partially reused, which causes vulnerable functions may not be in the target software \cite{OSSPolice}. As a result, existing methods tend to produce numerous false positives when their detected TPLs are directly associated with a vulnerability. LibAM is expected to solve this problem by reuse area detection techniques. We can extract the vulnerable function names from patches and match them with the reused function names in TPLs detected by the area matching framework. Out of the 2519 CVEs that we associated, 1240 had patches available. This enabled us to further filter the false positives. After the filtering process, we were able to reduce the number of vulnerabilities associated with TPL from 36,135 down to 14,863. To explain how LibAM filters incorrect vulnerabilities, we're presenting two examples. In the case of CVE-2014-9485, there's a vulnerability related to Path Traversal in the \textit{$do\_extract\_currentfile$} function within the \textit{minizip} library. While both \textit{precomp} and \textit{lzbench} use parts of \textit{minizip}'s code, they don't use the specific \textit{$do\_extract\_currentfile$} function. As a result, there will be two false positive vulnerabilities without area detection. LibAM handles this by identifying these reused areas and removing the false positives. Additionally, for CVE-2019-12900, there's a vulnerability involving an out-of-bounds write operation in the \textit{$BZ2\_decompress$} function in version 1.0.6 of the \textit{bzip2} library. Versions of \textit{minizip} from 2.0.0 to 2.7.5 use this vulnerable version of \textit{bzip2}, making them vulnerable to this particular issue. However, \textit{lzbench} isn't affected because it uses versions of \textit{bzip2} released after 1.8. Therefore, identifying the specific library versions also helps to eliminate incorrect vulnerability warnings.

Furthermore, by analyzing the results, we have the following findings, which demonstrate the benefits of area detection for discovering complex reuse relationships.

\begin{figure*}[htbp]
\begin{center}
\subfigure[CDF of the proportion of target binaries ]{
\includegraphics[width=2.4in]{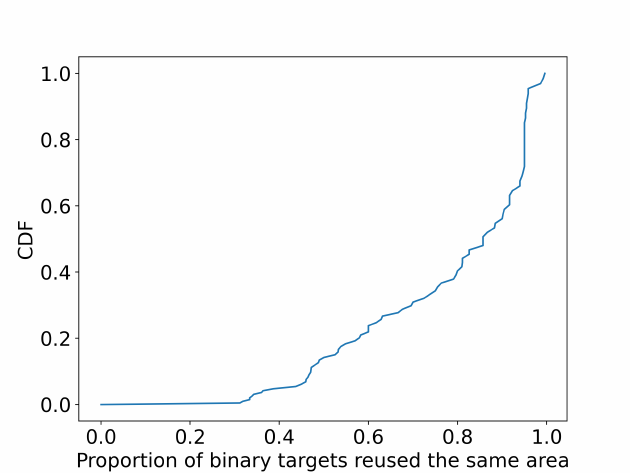}
}
\subfigure[CDF of the number of TPLs]{
\includegraphics[width=2.4in]{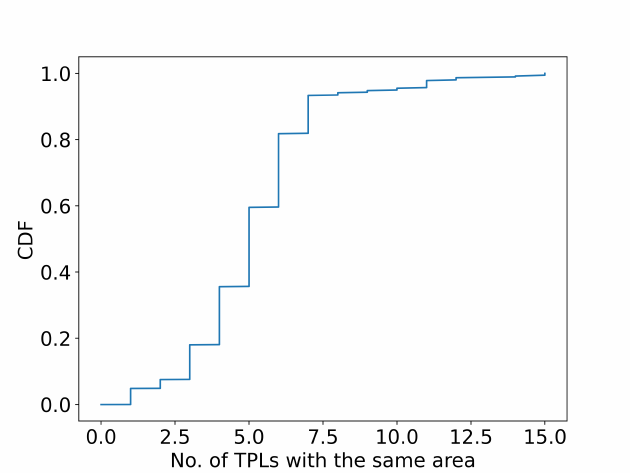}
}
\caption{CDF plots of the two findings. 12(a) indicates for each TPL, the proportion of target binaries reuse the same area of it. 12(b) indicates that for each target binary, the number of TPLs detected in the same area.}
\end{center}
\end{figure*}

\textbf{Different target binaries always tend to reuse the same area of TPL.}
In the detection results of Dataset\_FW, different areas of TPLs have different tendencies to be reused. In Figure 12(a), more than 85.8\% of TPLs meet the requirement of having the area reused by more than 50\% of the detected target binaries. Besides, 43.5\% of TPLs have more than 90\% of target binaries reusing the same area of it. This finding aligns with the ground truth from the public Dataset\_ISRD. For example, all four target binaries that reuse the \textit{Bzip2} share the same \textit{BZ2\_bzBufferToBuffCompress} and \textit{BZ2\_bzBufferToBuffDecompress} functions, as well as their respective subfunctions. This suggests that some areas of code in TPLs are more likely to be reused than others and that vulnerabilities in these code areas are more widespread in their impact. Consequently, when a code area from a TPL is identified as being reused, it is more probable that this area will also be reused by other software. Based on this finding, researchers can allocate additional manual analysis resources to scrutinize code areas that exhibit a higher likelihood of reuse, thus enhancing the effectiveness of vulnerability detection and mitigation efforts.

\textbf{There are numerous identical reuse areas in different TPLs.}
In Figure 12(b), more than 97.5\% of target binaries meet the requirement that the same area is detected to have reuse relationships with more than two TPLs. Moreover, The same area can be detected in up to 15 different TPLs. This is also reflected in the ground truth of Dataset\_ISRD. For instance, \textit{lzbench} reuses a large number of functions from \textit{precomp}, many of which are actually from \textit{Brotli}. Dataset\_ISRD only analyzes that \textit{lzbench} reuses \textit{precomp} and \textit{Brotli}, but after area detection, we find that these two reuses are actually the same area of code. This shows that area detection is able to detect complex reuse relationships. In addition, different areas of the TPLs may represent different functionalities. Distinguishing between areas that are reused by more TPLs and more specific areas belonging to only one TPL facilitates further manual analysis, e.g., analysis of areas that are reused more often will lead to more valuable results.

\begin{tcolorbox}[colback=gray!10,
                  colframe=black,
                  arc=1mm, auto outer arc,
                  boxrule=1.5pt,
                 ]
\textbf{Answering to RQ6:} LibAM can detect real-world TPLs on a large scale. We show one application of LibAM, vulnerability associating, by combining it with existing version identification works and vulnerability information. Besides, we make several interesting findings.

\end{tcolorbox}

\section{Discussion}

LibAM achieved high scores on both TPL detection tasks and Reuse area detection tasks. Although the accuracy of the reuse area detection task was not as high as that of the TPL detection task, we have proved that the Area Matching framework is feasible, and further research into a better Area Matching framework is worth pursuing. Moreover, we believe that Area Matching framework at the FCG level has research prospects for other applications of binary similarity calculation such as malware detection, software infringement, and patch analysis. 

Software reuse relationships are often complicated. We have addressed the issue of partial reuse through Area Matching framework. The detection at the area level makes the detection granularity consistent with the reuse granularity, greatly reducing the mistakes caused by existing approaches that set thresholds for file-level TPLs.

The existing works \cite{LibDB, ISRD, ModX}, including LibAM, are based on the results of function similarity calculation methods. However, even the state-of-the-art function similarity calculation methods may yield many false positives and false negatives in different optimization options and architectures \cite{jTrans}.  Although LibAM has significantly reduced this impact through Area Matching framework, the impact still exists. A better function similarity calculation method can further improve the accuracy of TPL detection.

LibAM focuses on the improvement of TPL detection at the area level compared to existing methods, leaving the improvement of the function similarity calculation work itself to help TPL detection for future research. We performed TPL detection for target binaries with different optimization options and architectures to demonstrate the stability of LibAM. We did not do other similar experiments across compilers, as in IoT firmware, we found that the main impact comes from the different compilation options, and the compilers tend to be only \textit{gcc}, with operating system mostly based on Linux. We leave the more diverse scenarios for future research.

Some callee functions may be called externally, such as using dynamic links. This can lead to differences in the target FCG and TPL FCG. However, by using the external dynamic link library as the target binary for LibAM, we can detect the missed TPLs caused by the FCG difference. In fact, in Dataset\_ISRD, all callee functions are within the binary. The only observed external call is the use of an additional springboard function for calls to the C standard library function in the x64 architecture,  which resulted in a minor difference from other architectures. Nevertheless, this difference did not have a significant impact on the detection process.

Some TPL detection works treat different versions of TPL as different libraries \cite{BAT, B2SFinder, ModX}, and other TPL detection works also do simple version identification by using the TPL detection technology \cite{OSSPolice, LibDB}. We believe that version identification using coarse-grained features at the library level is inaccurate because TPL detection aims to detect roughly similar code in a large amount of different code, while version identification aims to detect fine-grained differences in a large amount of the same code. We consider the TPL detection work as orthogonal to the version identification work. It is worth noting that LibAM can be directly combined with existing version identification work, as in Section 4.8, where we use the version identification method in OSSPolice \cite{OSSPolice}.

There is a bias between the list of potential vulnerabilities generated in Section 4.7 and the real vulnerabilities. Even though we have filtered out many false positives through version and reuse area location information, the target binary may patch known vulnerabilities or the target binary vulnerabilities cannot be triggered, which is a common problem faced by all TPL detection works. Vulnerabilities verifying work \cite{Fiber, Pdiff, QF_bugsearch, IR-Cross_arch_bug_search,patchScout, Palmtree, MVP} need to be done to further verify the quality of potential vulnerabilities. However, this paper has demonstrated the accuracy of TPL detection and reuse area detection compared with the existing methods, and vulnerability association is only one application of LibAM. We regard vulnerabilities verifying as orthogonal work for future research.

\section{Related Work}
\subsection{Binary Function Similarity Calculation}
Binary similarity calculation can be applied to various applications, such as reuse detection \cite{ISRD, LibDB}, malware identification \cite{BinSim}, vulnerability search \cite{QF_bugsearch, IR-Cross_arch_bug_search}, patches detection \cite{BinXray, Fiber, Pdiff}, etc. Recently, many researchers have focused on binary similarity calculation work.

During the compilation of C/C++ code, many changes occur and important information is lost, such as function names, variable names, source comments, data structure definitions, etc \cite{Code_Similarity_survey2}. Furthermore, the same source code can produce vastly different binaries when compiled with different options and architectures, making binary similarity calculation more difficult. A large number of works have been proposed for binary function similarity calculation. 

When discussing text hashing-based approaches like Gitz \cite{Gitz} and VIVA \cite{VIVA}, it is important to consider the inherent limitations of such techniques. While they offer simplicity and efficiency, text hashing methods can be sensitive to even minor changes in the binary code, potentially leading to false negatives in similarity detection. Furthermore, these methods may not be as effective when confronted with obfuscated code or code that has undergone significant transformations during compilation.

Symbolic execution-based techniques, such as BinSim \cite{BinSim} and Bingo \cite{Bingo}, offer a more robust approach to function similarity detection. By analyzing the possible execution paths and states of a binary function, these methods can provide a more comprehensive understanding of the underlying code. However, symbolic execution can be computationally expensive, particularly for large or complex codebases, and may be prone to path explosion and state space explosion issues.

Deep learning-based approaches \cite{fcs_cross_opti_obus, FSC_method1, QF_FCS_ML} have gained traction in recent years due to their ability to automatically learn feature representations from raw data. These methods can effectively capture complex patterns and relationships in the binary code, potentially leading to more accurate similarity detection. Nevertheless, deep learning models can be resource-intensive, requiring significant computational power and training data to achieve optimal accuracy.

The use of NLP techniques, such as JTrans \cite{jTrans}, in binary function similarity calculation represents an innovative approach that leverages the advancements made in natural language processing. By treating binary code as a form of language, these methods can apply well-established NLP techniques to analyze and compare binary functions. However, such approaches may still face challenges when dealing with the diverse nature of binary code, which can differ significantly from natural language in terms of structure and semantics.

Incorporating multidimensional features, as suggested by recent research \cite{Ordermatters, Vulhawk}, is an essential step towards improving function similarity matching. By considering features like strings, control flow graphs, and data flow information, these methods can provide a more holistic view of the binary code, leading to more accurate and reliable similarity detection. Future research could explore the integration of various features and their respective weights, enabling more fine-tuned and adaptable similarity calculation models.


However, with the increase of candidate functions and the emergence of complicated problems such as function inlining, the accuracy of existing function similarity matching methods is seriously degraded, and new methods need to be proposed to address this problem. The purpose of this paper is to use the results of the function similarity computation as anchors to evaluate the improvement brought by the area-level similarity computation for the TPL detection task, without focusing on the improvement of the function similarity computation task itself.

\subsection{Vulnerability Detection}
Vulnerability detection has long been a popular and significant field in computer science and cybersecurity. Researchers aim to detect vulnerabilities in newly developed code by extracting features from vulnerable functions and determining if these vulnerable functions exist within the target code. This process typically involves calculating the similarity between functions, which forms the basis for most vulnerability detection methodologies.

Early works in this field, such as Bingo \cite{Bingo} and SAFE \cite{SAFE}, calculated function similarity by directly comparing the vulnerable function with all functions present in the target code. The resulting function similarity score was then utilized as the vulnerability detection result. However, these pure function similarity calculations were unable to capture fine-grained features of vulnerabilities, such as whether the vulnerabilities were patched or not. Consequently, later research began to incorporate patch code information to enhance vulnerability detection.

More recent approaches, including MVP \cite{MVP} and VIVA \cite{VIVA}, detect vulnerabilities by employing data stream slicing techniques to identify the presence of vulnerability code and the absence of patch code. Additionally, Fiber \cite{Fiber} and PDiff \cite{Pdiff} attempt to extract deep patch code semantics by utilizing symbolic execution to ascertain whether the target vulnerable function has been patched or not.

Both Third-Party Library (TPL) detection and vulnerability detection work can detect vulnerabilities due to code reuse or similarity. Their differences lie in the following aspects: Vulnerability detection methodologies are typically designed to identify vulnerabilities on a small, fine-grained scale, often incorporating patch information to enhance detection results. In contrast, TPL detection approaches aim to detect code reuse on a larger scale and can improve detection results of individual vulnerable functions through global reuse information.

Moreover, TPL detection is not limited to correlating one-day vulnerabilities but can also analyze software components, detect software plagiarism, correlate malware, and more. Combining patch information from vulnerability detection methodologies with the global information from TPL detection approaches may contribute to a more comprehensive and effective vulnerability detection framework.

By integrating the strengths of both vulnerability detection and TPL detection techniques, researchers can create more robust and accurate systems for identifying and addressing potential security risks in software development. This holistic approach will ultimately contribute to enhancing overall cybersecurity and ensuring the reliability of software systems in various domains.


\subsection{Third-Party Libraries Detection}
Third-party libraries (TPLs) are essential components of modern software development, as they provide ready-made functionality and enable developers to focus on the core aspects of their applications. However, the introduction of insecure third-party libraries can bring new threats to software, making the effective detection of reused third-party libraries in target software a hot research topic.

Previous works  \cite{java1,java2,java3,java4,java5,java7,ORLIS} have focused on TPL detection tasks in Java, as the Java package allows for the easy extraction of pseudocode. Consequently, these works on Java code reuse have proven to be effective. However, C/C++ binaries face more significant challenges in TPL detection, as much information is stripped and code changes greatly depending on the optimization options and architectures used.

Recently, several works \cite{OSSPolice,ISRD,LibDB,LibDX} have been proposed to address TPL detection in C/C++ binaries. Some of these works, such as BAT \cite{BAT}, OSSPolice \cite{OSSPolice}, and B2SFinder \cite{B2SFinder}, leverage constant features as fingerprints to detect TPL. However, they exhibit poor recall in binaries that lack sufficient constant features. Other works, including ISRD \cite{ISRD} and LibDB \cite{LibDB}, attempt to use function similarity calculation technology to detect TPL reuse. CENTRIS \cite{CENTRIS}, for example, employs the Trending Local Sensitive Hash Algorithm (TLSH) for function similarity calculation to detect TPL reuse. However, due to the impact of different compilation environments, the results of directly applying function similarity calculation to TPL detection are unsatisfactory \cite{jTrans}.

Existing methods often attempt to reduce false positives and false negatives through further filtering. For instance, ISRD \cite{ISRD} considers reuse by identifying more than half of the function similarity, while LibDB \cite{LibDB} relies on more than three connected functions on Function Call Graphs (FCGs) to judge reuse. Nevertheless, these filtering methods are based on isolation function matching, and the final results still contain numerous false negatives and false positives. Furthermore, these methods demonstrate poor accuracy under different optimization options and architectures.

However, these filtering methods are simple and the final results are still many false negatives and false positives. They all compare isolated function similarities and filter them with some rules that lead to poor accuracy under different optimization options and architectures. We try to explore reuse areas on FCG and conduct area matching rather than isolation function matching to obtain both high precision and recall.

\section{Conclusion}
In this paper, we proposed LibAM, a novel Area Matching framework to transform the TPL detection task into the TPL reuse area matching task, so as to obtain both high accuracy and robustness. Meanwhile, we detect specific reuse areas, thus providing interpretable evidence for TPL detection results and helping to detect complex reuse relationships and downstream tasks.

LibAM stands out from previous methods by overcoming the challenges of exactly detecting  TPLs and identifying exact reuse areas efficiently. Our method demonstrates a marked improvement over state-of-the-art (SOTA) techniques by detecting exact TPLs and achieving ideal results in identifying reuse areas. To validate the effectiveness and efficiency of LibAM, we conducted extensive experiments across various optimization options and architectural frameworks. The results consistently showed that LibAM outperforms existing SOTA work in TPL detection tasks, making it a highly desirable solution for a wide range of applications. Furthermore, we evaluated the accuracy of LibAM in large-scale and real-world binaries extracted from Internet of Things (IoT) firmware to investigate its practical applicability. By doing so, we were able to generate a potential vulnerability list, which can prove invaluable for researchers and practitioners working on IoT security.

In conclusion, LibAM represents a significant advancement in the field of TPL detection, offering a comprehensive and efficient solution for detecting exact areas and identifying reuse areas. In future work, we plan to extend LibAM to overcome other security challenges in the software and hardware domain, further exploring its wide-ranging applicability and impact.



\section*{Acknowledgements}
We thank the associated editor and anonymous reviewers of TOSEM for their valuable feedback. This work was partially supported by the National Key Research and Development Program of China (2022YFB3103904), the National Natural Science Youth Foundation (62002342), and the National Natural Science Foundation of China (61931019).

\bibliography{ref}










\end{document}